\documentclass[preprint,aps]{revtex4}

\usepackage{graphicx}
\usepackage{lscape}

\begin{document}

\def\ds{\displaystyle}
\def\beq{\begin{equation}}
\def\eeq{\end{equation}}
\def\bea{\begin{eqnarray}}
\def\eea{\end{eqnarray}}
\def\beeq{\begin{eqnarray}}
\def\eeeq{\end{eqnarray}}
\def\ve{\vert}
\def\vel{\left|}
\def\ver{\right|}
\def\nnb{\nonumber}
\def\ga{\left(}
\def\dr{\right)}
\def\aga{\left\{}
\def\adr{\right\}}
\def\lla{\left<}
\def\rra{\right>}
\def\rar{\rightarrow}
\def\nnb{\nonumber}
\def\la{\langle}
\def\ra{\rangle}
\def\ba{\begin{array}}
\def\ea{\end{array}}
\def\tr{\mbox{Tr}}
\def\ssp{{\Sigma^{*+}}}
\def\sso{{\Sigma^{*0}}}
\def\ssm{{\Sigma^{*-}}}
\def\xis0{{\Xi^{*0}}}
\def\xism{{\Xi^{*-}}}
\def\qs{\la \bar s s \ra}
\def\qu{\la \bar u u \ra}
\def\qd{\la \bar d d \ra}
\def\qq{\la \bar q q \ra}
\def\gGgG{\la g^2 G^2 \ra}
\def\q{\gamma_5 \not\!q}
\def\x{\gamma_5 \not\!x}
\def\g5{\gamma_5}
\def\sb{S_Q^{cf}}
\def\sd{S_d^{be}}
\def\su{S_u^{ad}}
\def\ss{S_s^{??}}
\def\ll{\Lambda}
\def\lb{\Lambda_b}
\def\sbp{{S}_Q^{'cf}}
\def\sdp{{S}_d^{'be}}
\def\sup{{S}_u^{'ad}}
\def\ssp{{S}_s^{'??}}
\def\sig{\sigma_{\mu \nu} \gamma_5 p^\mu q^\nu}
\def\fo{f_0(\frac{s_0}{M^2})}
\def\ffi{f_1(\frac{s_0}{M^2})}
\def\fii{f_2(\frac{s_0}{M^2})}
\def\O{{\cal O}}
\def\sl{{\Sigma^0 \Lambda}}
\def\es{\!\!\! &=& \!\!\!}
\def\ar{&+& \!\!\!}
\def\ek{&-& \!\!\!}
\def\cp{&\times& \!\!\!}
\def\se{\!\!\! &\simeq& \!\!\!}
\def\hml{\hat{m}_{\ell}}
\def\rr{\hat{r}_{\Lambda}}
\def\ss{\hat{s}}

\def\simlt{\stackrel{<}{{}_\sim}}
\def\simgt{\stackrel{>}{{}_\sim}}


\title{
         {\Large
                 {\bf
Double-Lepton Polarization Asymmetries in  $B_s \rar \phi \ell^+
\ell^-$ Decay in the Fourth-Generation Standard Model
                 }
         }
      }

\author{S. M. Zebarjad\footnote{zebarjad@physics.susc.ac.ir}, F. Falahati, H. Mehranfar}
\affiliation{Physics Department,  Shiraz University, Shiraz 71454,
Iran}

\begin{abstract}
 In this paper, we investigate the effects of the fourth
generation of quarks on the  double-lepton polarization
asymmetries in the $B_s \rar \phi \ell^+ \ell^-$ decay. It is
shown that these asymmetries in $B_s \rar \phi \ell^+ \ell^-$
decay compared with those of $B \rar K \ell^+ \ell^-$ decay are
more sensitive to the fourth-generation parameters.  We conclude
that an efficient way to establish the existence of the fourth
generation of quarks could be  the study of these asymmetries in
the $B_s \rar \phi \ell^+ \ell^-$ decay.

\end{abstract}

\maketitle

\section{Introduction}
Although  Standard Model (SM) is a  successful theory, there is no
clear theoretical argument within this model to restrict the
number of generations to three, and therefore the possibility of a
new generation should not be ruled out.
 Based on this possibility, a
number of theoretical and experimental investigations have been
performed.
 The measurement of the $Z$ decay widths restricts the number of light
 neutrino for $m_{\nu}<m_Z/2$ to three\cite{ewwg}. However, if a heavy
 neutrino exits, the possibility of extra generations of heavy quarks is not
 excluded from the experiment. Moreover the electro weak data \cite{okun} supports an extra
 generation of heavy quarks, if the mass difference between the new up and down-type quarks is not
 too large.

Many authors who support the existence of fourth-generation
studied those effects in various areas, for instance Higgs and
neutrino physics, cosmology and dark matter
\cite{Polonsky}--\cite{Ginzburg}. For example, in \cite{Ginzburg}
it is argued that the fourth generation of quarks and leptons can
be generated in the Higgs boson production at the Tevatron and the
LHC, before being actually detected.
 By the detailed study of this process at the Tevatron and LHC,
the number of generations in the SM can be determined. Moreover,
the flavor democracy (Democratic Mass Matrix approach) \cite{FD1}
favors the existence of the nearly degenerate fourth SM family,
while the fifth SM family is disfavored both by the mass
phenomenology and precision tests of the ${\rm SM}$ \cite{FD2}.
The main restrictions on the new SM families come from the
experimental data on the $\rho$ and $S$ parameters \cite{FD2}.
However, the common mass of the fourth quark ($m_{t'}$) lies
between 320 $GeV$ and 730 $GeV$ considering the experimental value
of $\rho=1.0002^{+0.0007}_{-0.0004}$ \cite{PDG}. The last value is
close to upper limit on heavy quark masses, $m_q\leq 700$ $GeV$
$\approx 4m_t$, which follows from partial-wave unitarity at high
energies \cite{chanowitz}. It should be noted that with preferable
value $a\approx g_w$ Flavor Democracy predicts $m_{t'}\approx 8
m_w\approx 640$ $GeV$.

One of the promising areas in the experimental search for the
fourth-generation, via its indirect loop effects, is the rare B
meson decays. Based on this idea, serious attempts to probe the
effects of the fourth-generation
 on the rare B meson were made by many researchers.
The fourth-generation can affect physical observables, i.e.
branching ratio, CP asymmetry, polarization asymmetries and
forward--backward asymmetries. The study of these physical
observables is a good tool to look for the fourth generation of up
type quarks \cite{Hou:2006jy}--\cite{Turan:2005pf}.

Recently,  the sensitivity of the double-lepton polarization
asymmetries to the fourth-generation in the transition of $B$ to a
pseudo scalar meson  ($B \rightarrow K\ell^{+}\ell^{-}$) has been
investigated and it is found out that this observable is sensitive
to the fourth-generation parameters ($m_{t'}$,
$V_{t'b}V^*_{t's}$)\cite{Bashiry:2007tf}. In this work, we
investigate the effects of the fourth generation of quarks
$(b',t')$ on the double-lepton polarizations in  the transition of
$B$ to a vector meson
 ( $B_s\rightarrow \phi \ell^{+}\ell^{-}$) and compare
our results with those of $B \rightarrow K\ell^{+}\ell^{-}$ decay
presented in Ref.\cite{Bashiry:2007tf}. It should be mentioned
that both decays occur through $b\rightarrow s$ transition in
which
 the sequential fourth generation of up
quarks $(t^{\prime})$, like $ u,c,t$ quarks, contributes at the
loop level. Hence, this new generation will change only the values
of the Wilson coefficients via virtual exchange of the
fourth-generation up quark $t^{\prime}$ and the full operator set
is exactly the same as in SM.

The paper is organized as follows. In Section II, the expressions
for the matrix element and double-lepton polarizations of $B_s
\rightarrow \phi \ell^+ \ell^-$ in the SM have been presented. The
effect of the fourth generation of quarks on the effective
Hamiltonian and the double-lepton polarization asymmetries have
been discussed in Section III. The sensitivity of these
polarizations to the fourth-generation parameters
$(m_{t'},r_{sb},\phi_{sb})$ have been numerically analyzed in the
final Section.

\section{The Matrix Element and Double-Lepton Polarizations     of $B_s
\rightarrow \phi \ell^+ \ell^-$ in the SM}\label{SM}

In the SM, the relevant  effective Hamiltonian for $B_s \rightarrow
\phi \ell^+ \ell^-$ decay which is described by $b \rightarrow s
\ell^+ \ell^-$ transition at quark level can be written as
\begin{equation}
{\cal H}_{\rm eff} = -\frac{ G_F}{\sqrt{2}} V_{tb} V^*_{ts}
\sum_{i=1}^{10} C_i(\mu) {\cal O}_i(\mu) \,,
\end{equation}
where the complete set of the operators ${\cal O}_i(\mu)$ and the
corresponding expressions for the Wilson coefficients $C_i(\mu)$
are given in \cite{R23}. Using the above effective Hamiltonian,
the one-loop matrix elements of $b \rightarrow s \ell^+ \ell^-$
can be written in terms of the tree-level matrix elements of the
effective operators as:
\begin{eqnarray}\label{bdecay}{\cal M}(b \rightarrow s \ell^+ \ell^-)&=& <s \ell^+  \ell^- |{\cal
H}_{\rm eff}|b> \nonumber\\
&= &-\frac{ G_F}{\sqrt{2}} V_{tb} V^*_{ts} \sum_i C^{\rm eff}_i(\mu)
<s \ell^+\ell^-|{\cal O}_i|b>^{tree}.\nonumber\\
 &=&-\frac{G_{F}
\alpha}{2\pi\sqrt{2} }V_{tb}V^{*}_{ts} \Bigg[\tilde{C}_{9}^{\rm
eff}\bar{s}\gamma_{\mu}(1-\gamma_{5})b~ \bar{\ell}\gamma_{\mu}
\ell+\tilde{C}_{10}^{\rm eff}\bar{s}\gamma_{\mu}(1-\gamma_{5})b~
\bar{\ell}\gamma_{\mu} \gamma_{5}\ell
\nonumber\\&&\,\,\,\,\,\,~~~~~~~~~~~~~- 2 C_7^{\rm
eff}\frac{m_b}{q^2} \bar s \sigma_{\mu\nu} q^\nu (1+\gamma_5) b \,
\bar \ell \gamma_\mu \ell \Bigg],
\end{eqnarray}
where $q^2=(p_1+p_2)^2$ and $p_1$ and $p_2$ are the final leptons
four--momenta and the effective Wilson coefficients at $\mu$
scale, are given as \cite{R23,R24}:
\begin{eqnarray}\label{c9}
C_7^{\rm eff}&=&C_7-\frac{1}{3}C_5-C_6 \nonumber\\
C_{10}^{\rm eff} &=&\frac{\alpha}{2\pi}\tilde{C}^{\rm eff}_{10}=C_{10}\nonumber\\
C_{9}^{\rm eff} &=&\frac{\alpha}{2\pi}\tilde{C}^{\rm
eff}_{9}=C_9+\frac{\alpha}{2\pi}Y(s).
\end{eqnarray}
In Eq.(\ref{c9}),  $s = q^2 / m_b^2$ and the function $Y(s)$
contains the short-distance contributions due to the one-loop
matrix element of the four quark operators, $Y_{per}(s)$,  as well
as the long-distance contributions coming from the real $c\bar c$
intermediate states, i.e., $J/\psi$, $\psi^\prime$, $\cdots$.The
latter contributions are taken into account by introducing
Breit--Wigner form of the resonance propagator which leads to the
second term in the following formula (see Eq.\ref{Yper})
\cite{R26}--\cite{R28}.
 As a result the function $Y(s)$ can be written as:
 \begin{eqnarray}\label{Yper}
  Y(s) &=& Y_{per}(s) +
\frac{3\pi}{\alpha^2}  (3 C_1 + C_2 + 3 C_3 + C_4 + 3 C_5 + C_6)\\
\nonumber &\times&\sum_{V_i=\psi_i} \kappa_i \frac{m_{V_i}
\Gamma(V_i \rightarrow \ell^+ \ell^-)} {m_{V_i}^2 - s m_b^2 - i
m_{V_i} \Gamma_{V_i}},
\end{eqnarray} where
 \begin{eqnarray} Y_{per}(s) &=&
g(\frac{m_c}{m_b}, s) (3 C_1 + C_2 + 3 C_3 + C_4 + 3 C_5 + C_6)
\nonumber
\\&-& \frac{1}{2} g(1, s) (4 C_3 + 4 C_4 + 3 C_5 + C_6) \nonumber
\\&-& \frac{1}{2} g(0, s) (C_3 + 3 C_4) + \frac{2}{9} (3 C_3 +
C_4 + 3 C_5 + C_6). \end{eqnarray} The  explicit expressions for
the $g$ functions can be found  in \cite{R23} and the
phenomenological parameters $\kappa_i$ in Eq.(\ref{Yper}) can be
determined from
\begin{equation} {\cal
B} (B \rightarrow K^\ast V_i \rightarrow K^\ast \ell^+ \ell^-)={\cal
B} (B \rightarrow K^\ast V_i)\, {\cal B} ( V_i \rightarrow \ell^+
\ell^-), \end{equation}
 where the data for the right hand side is
given in \cite{R29}. For the lowest resonances, $J/\psi$ and
$\psi^\prime$ one can use $\kappa = 1.65$ and $\kappa = 2.36$,
respectively (see \cite{R30}). In this study, we neglect the
long-distance contributions for simplicity and like
Ref.\cite{R23}, to have a scheme independent matrix element, we
use the leading order  as well as the next-to-leading order QCD
corrections to $C_9$ and the leading order QCD corrections to
 the other Wilson coefficients.
\par  In order to compute the decay width and other physical
observables of $B_s \rightarrow \phi \ell^+ \ell^-$ decay, we need
to sandwich the matrix elements in  Eq.(\ref{bdecay}) between the
final and initial meson states. Therefore, the hadronic matrix
elements for the $B_s \rightarrow \phi \ell^+ \ell^-$ can be
parameterized in terms of  form factors.  For the vector meson
$\phi$ with polarization vector $\varepsilon_\mu$ the semileptonic
form factors of the V--A current is defined as:
\begin{eqnarray}\label{form1}
<\phi(p_{\phi},\epsilon) \mid \bar{s}\gamma_{\mu}(1-\gamma_{5})b\mid
B(p_{B_s})> = -\frac{2V(q^{2})}{m_{B_s}+m_{\phi}}
\epsilon_{\mu\nu\rho\sigma}p_{\phi}^{\rho}q^{\sigma}\epsilon^{*\nu}\nonumber
\\
 -i \left[\epsilon_{\mu}^{*}(m_{B_s}+m_{\phi})A_{1}(q^{2})
-(\epsilon^{*}q)(p_{B_s}+p_{\phi})_{\mu}\frac{A_{2}(q^{2})}{m_{B_s}+m_{\phi}}
\right. \nonumber \\
 - \left. q_{\mu}(\epsilon^{*}q)\frac{2m_{\phi}}{q^{2}}
(A_{3}(q^{2})-A_{0}(q^{2})) \right],
\end{eqnarray}
where $q=p_{B_s}-p_{\phi}$, and $A_{3}(q^{2}=0)=A_{0}(q^{2}=0)$
(this condition ensures that there is no kinematical singularity
in the matrix element at $q^2=0$). Also, the form factor
$A_{3}(q^{2})$ can be written as a linear combination of the form
factors $A_{1}$ and $A_{2}$ :
\begin{eqnarray}\label{form2}
A_{3}(q^{2})=\frac{1}{2m_{\phi}}\left[(m_{B_s}+m_{\phi})A_{1}(q^{2})-
(m_{B_s}-m_{\phi})A_{2}(q^{2})\right].
\end{eqnarray}
The  other semileptonic form factors coming from the dipole operator
$\sigma_{\mu\nu} q^\nu (1 + \gamma_5) b$ can be defined as:
\begin{eqnarray} \label{form3}\lefteqn{ \lla \phi(p_{\phi},\varepsilon) \vel \bar s
i \sigma_{\mu\nu} q^\nu (1 + \gamma_5) b \ver B(p_{B_s}) \rra =}
\nnb \\&&4 \epsilon_{\mu\nu\rho\sigma} \varepsilon^{\ast\nu} p^\rho
q^\sigma T_1(q^2) + 2 i \left[ \varepsilon_\mu^\ast
(m_{B_s}^2-m_{\phi}^2) -
(p_{B_s} + p_{\phi})_\mu (\varepsilon^\ast q) \right] T_2(q^2) \nnb \\
&&+ 2 i (\varepsilon^\ast q) \left[ q_\mu - (p_{B_s} +
p_{\phi})_\mu \frac{q^2}{m_{B_s}^2-m_{\phi}^2} \right] T_3(q^2)~.
\end{eqnarray} As seen From Eqs. (\ref{form1}-\ref{form3}),  we have to compute
the form factors to obtain  the physical observables at hadronic
level.The form factors are related  to the non-perturbative sector
of QCD and can be evaluated only by using non-perturbative
methods. In the present work, we use light cone QCD sum rule
predictions for the form factors
   in which one-loop radiative
corrections to twist-2 and twist-3 contributions are taken into
account. The form factors  \bea F(q^{2})\in\{V(q^2), A_{0}(q^{2}),
A_1(q^{2}), A_{2}(q^{2}), A_{3}(q^{2}), T_{1}(q^{2}), T_{2}(q^{2}),
T_{3}(q^{2})\}~,\nnb \eea
 are  fitted  to the the following functions \cite{R31,R32}:
\begin{equation}
F(q^2)=\frac{F(0)}{1-a_F \frac{q^2}{m_{B_s}^2}+b_F
(\frac{q^2}{m_{B_s}^2})^2},
\end{equation}
where the parameters $F(0)$, $a_F$ and $b_F$ are listed in the
Table\ref{tt}.
\begin{table}[h]
\renewcommand{\arraystretch}{1.5}
\addtolength{\arraycolsep}{3pt}
$$
\begin{array}{|l|ccc|}
\hline & F(0) & a_F & b_F \\ \hline
A_0^{B_s \rar \phi} &\phantom{-}0.382 & 1.77 & \phantom{-} 0.856 \\
A_1^{B_s \rar \phi} &\phantom{-}0.296 & 0.87 & -0.061 \\
A_2^{B_s \rar \phi} &\phantom{-}0.255 & 1.55 & \phantom{-} 0.513\\
V^{B_s \rar \phi}   &\phantom{-}0.433 & 1.75 & \phantom{-} 0.736\\
T_1^{B_s \rar \phi} &\phantom{-}0.174 & 1.82 & \phantom{-} 0.825\\
T_2^{B_s \rar \phi} &\phantom{-}0.174 & 0.70 & -0.315\\
T_3^{B_s \rar \phi} &\phantom{-}0.125 & 1.52 & \phantom{-} 0.377\\
\hline
\end{array}
$$
\caption{The form factors for $B_s\rightarrow \phi
\,\ell^{+}\ell^{-}$ in a three--parameter fit
\cite{R31}.}\label{tt}
\renewcommand{\arraystretch}{1}
\addtolength{\arraycolsep}{-3pt}
\end{table}

 Using
 Eqs.(\ref{form1}-\ref{form3}),  the matrix element of the $B_s \rar \phi
\ell^+ \ell^-$ decay can be written as follows: \bea \lefteqn{
\label{e6306} {\cal M}(B_s\rightarrow \phi \ell^{+}\ell^{-}) =
\frac{G \alpha}{4 \sqrt{2} \pi} V_{tb} V_{ts}^\ast }\\
&&\times \Bigg\{ \bar \ell \gamma^\mu(1-\gamma_5) \ell \, \Big[ -2
B_0 \epsilon_{\mu\nu\lambda\sigma} \varepsilon^{\ast\nu}
p_{\phi}^\lambda q^\sigma
 -i B_1 \varepsilon_\mu^\ast
\nnb \\
&&~~~~~~~~~~~~~~~~~~~~~~~~+ i B_2 (\varepsilon^\ast q) (p_{B_s}+p_{\phi})_\mu + i B_3 (\varepsilon^\ast q) q_\mu  \Big] \nnb \\
&&~~~~+ \bar \ell \gamma^\mu(1+\gamma_5) \ell \, \Big[ -2 C_1
\epsilon_{\mu\nu\lambda\sigma} \varepsilon^{\ast\nu}
p_{\phi}^\lambda q^\sigma
 -i D_1 \varepsilon_\mu^\ast
\nnb \\
 &&~~~~~~~~~~~~~~~~~~~~~~~~~~+ i
D_2 (\varepsilon^\ast q) (p_{B_s}+p_{\phi})_\mu + i D_3
(\varepsilon^\ast q) q_\mu  \Big] \Bigg\}~,\nnb \eea where \bea
\label{e6307} B_0 &=& (\tilde{C}^{\rm eff}_{9}- \tilde{C}^{\rm
eff}_{10}) \frac{V}{m_{B_s}+m_{\phi}} + 4 (m_{B_s}+m_s){C}^{\rm
eff}_{7} \frac{T_1}{q^2} ~, \nnb \\
B_1 &=& (\tilde{C}^{\rm eff}_{9}- \tilde{C}^{\rm eff}_{10})
(m_{B_s}+m_{\phi}) A_1 + 4 (m_{B_s}-m_s){C}^{\rm eff}_{7}
(m_{B_s}^2-m_{\phi}^2)
\frac{T_2}{q^2} ~, \nnb \\
B_2 &=& \frac{\tilde{C}^{\rm eff}_{9}- \tilde{C}^{\rm
eff}_{10}}{m_{B_s}+m_{\phi}} A_2 + 4 (m_{B_s}-m_s){C}^{\rm
eff}_{7} \frac{1}{q^2}  \left[ T_2 +
\frac{q^2}{m_{B_s}^2-m_{\phi}^2} T_3 \right]~,
\nnb \\
B_3 &=& 2 (\tilde{C}^{\rm eff}_{9}- \tilde{C}^{\rm eff}_{10})
m_{\phi} \frac{A_3-A_0}{q^2}-4
 (m_{B_s}-m_s){C}^{\rm
eff}_{7} \frac{T_3}{q^2} ~, \nnb \\
C_1 &=& B_0 ( \tilde{C}^{\rm eff}_{10} \rar -\tilde{C}^{\rm eff}_{10}\nnb)~, \\
D_i &=& B_i ( \tilde{C}^{\rm eff}_{10} \rar -\tilde{C}^{\rm
eff}_{10}\nnb)~,~~~~(i=1,~2,~3). \eea
 From the above equations for
the differential decay width, we get the following result: \bea
\label{e6308} \frac{d\Gamma}{d\hat{s}}(B_s \rar \phi \ell^+
\ell^-) = \frac{G^2 \alpha^2 m_{B_s}}{2^{14} \pi^5} \vel
V_{tb}V_{ts}^\ast \ver^2 \lambda^{1/2}(1,\hat{r},\hat{s}) v
\Delta(\hat{s})~, \eea with \bea \label{e6309}
\Delta&=&\frac{2}{3\hat{r}_{\phi}\hat{s}}m_{B_s}^2Re[
-12m_{B_s}^2\hat{m_{l}}^2\lambda\hat{s}\{(B_{3}-D_{2}-D_{3})B_{1}^{*}-(B_{3}+B_{2}-D_{3})D_{1}^{*}\}\nnb\\
\nonumber&&+ 12m_{B_s}^4\hat{m_{l}}^2\lambda\hat{s}(1-\hat{r}_{\phi})(B_{2}-D_{2})(B_{3}^{*}-D_{3}^{*})\\
\nonumber&&+
48\hat{m_{l}}^2\hat{r}_{\phi}\hat{s}(3B_{1}D_{1}^{*}+2m_{B_s}^4\lambda
B_0C_{1}^{*})\\
\nonumber &&-16m_{B_s}^4\hat{r}_{\phi}\hat{s}\lambda(\hat{m_{l}}^2-\hat{s})\{|B_0|^{2}+|C_{1}|^{2}\}\\
\nonumber &&-6m_{B_s}^4\hat{m_{l}}^2\lambda\hat{s}\{2(2+2\hat{r}_{\phi}-\hat{s})B_{2}D_{2}^{*}-\hat{s}|(B_{3}-D_{3})|^{2}\}\\
\nonumber &&-4m_{B_s}^2\lambda\{\hat{m_{l}}^2(2-2\hat{r}_{\phi}+\hat{s})+\hat{s}(1-\hat{r}_{\phi}-\hat{s})\}(B_{1}B_{2}^{*}+D_{1}D_{2}^{*})\\
\nonumber &&+\hat{s}\{6\hat{r}_{\phi}\hat{s}(3+v^2)+\lambda(3-v^2)\}\{|B_{1}|^{2}+|D_{1}|^{2}\}\\
&&-2m_{B_s}^4\lambda\{\hat{m_{l}}^2[\lambda-3(1-\hat{r}_{\phi})^2]-\lambda\hat{s}\}\{|B_{2}|^{2}+|D_{2}|^{2}\}],
\nonumber\eea where $\hat{s}=q^2/m_{B_s}^2$,
$\hat{r}_{\phi}=m_{\phi}^2/m_{B_s}^2$ and
$\lambda(a,b,c)=a^2+b^2+c^2-2ab-2ac-2bc$,
$\hat{m}_\ell=m_\ell/m_{B_s}$, $v=\sqrt{1-4\hat{m}_\ell^2/\hat{s}}$
is the final lepton velocity.

Having obtained the matrix element for the $B_s \rightarrow \phi
\ell^+ \ell^-$,  we can now calculate the double--polarization
asymmetries. For this purpose,  we define the orthogonal unit
vectors $s^{\pm \mu}_i$ in the rest frame of leptons, where i=L,N
or T refer to the longitudinal, normal and transversal
polarization directions, respectively:
 \bea \label{e6010} s^{-\mu}_L
\es \ga 0,\vec{e}_L^{\,-}\dr = \ga
0,\frac{\vec{p}_-}{\vel\vec{p}_-
\ver}\dr,~~~~~~~~~~~~~~~~~s^{+\mu}_L = \ga 0,\vec{e}_L^{\,+}\dr =
\ga 0,\frac{\vec{p}_+}{\vel\vec{p}_+ \ver}\dr, \nnb \\
s^{-\mu}_N \es \ga 0,\vec{e}_N^{\,-}\dr = \ga
0,\frac{\vec{p}_{\phi}\times \vec{p}_-}{\vel \vec{p}_{\phi}\times
\vec{p}_- \ver}\dr,~~~~~~~~~~s^{+\mu}_N = \ga 0,\vec{e}_N^{\,+}\dr
= \ga 0,\frac{\vec{p}_{\phi}\times
\vec{p}_+}{\vel \vec{p}_{\phi}\times \vec{p}_+ \ver}\dr, \nnb \\
s^{-\mu}_T \es \ga 0,\vec{e}_T^{\,-}\dr = \ga 0,\vec{e}_N^{\,-}
\times \vec{e}_L^{\,-} \dr,~~~~~~~~~~~~s^{+\mu}_T = \ga
0,\vec{e}_T^{\,+}\dr = \ga 0,\vec{e}_N^{\,+}
\times \vec{e}_L^{\,+}\dr. \nnb \\
 \eea
 In the above equations  $\vec{p}_\mp$ and
$\vec{p}_{\phi}$ are the three--momenta of the leptons $\ell^\mp$
and $\phi$ meson, respectively. Then by Lorentz transformation
these unit vectors are boosted from the rest frame of leptons to
the center of mass (CM) frame of leptons. Under this
transformation only the longitudinal unit vectors $s^{\pm \mu}_L $
change, but the other two vectors remain unchanged. $s^{\pm \mu}_L
$ in the CM frame of leptons are obtained as:
 \bea \label{e6011} \ga
s^{-\mu}_L \dr_{CM} = \ga \frac{\vel\vec{p}_- \ver}{m_\ell}, \frac{E
\vec{p}_-}{m_\ell \vel\vec{p}_- \ver}\dr,~~~~~~~~ \ga s^{+\mu}_L
\dr_{CM} = \ga \frac{\vel\vec{p}_- \ver}{m_\ell}, -\frac{E
\vec{p}_-}{m_\ell \vel\vec{p}_- \ver}\dr. \eea The polarization
asymmetries can now be calculated using the spin projector ${1 \over
2}(1 + \gamma_5 \!\!\not\!\! s_i^{-})$ for $\ell^-$ and the spin
projector ${1 \over 2}(1 + \gamma_5\! \not\!\! s_i^{+})$ for
$\ell^+$.

Considering the above explanations, we can  define the
double--lepton polarization asymmetries as in \cite{Fukae}: \bea
\label{doubleLepton} P_{ij}(\hat{s}) = \frac{\ds{\Bigg(
\frac{d\Gamma}{d\hat{s}}(\vec{s}_i^-,\vec{s}_j^+)}-
\ds{\frac{d\Gamma}{d\hat{s}}(-\vec{s}_i^-,\vec{s}_j^+) \Bigg)} -
\ds{\Bigg( \frac{d\Gamma}{d\hat{s}}(\vec{s}_i^-,-\vec{s}_j^+)} -
\ds{\frac{d\Gamma}{d\hat{s}}(-\vec{s}_i^-,-\vec{s}_j^+)\Bigg)}}
{\ds{\Bigg( \frac{d\Gamma}{d\hat{s}}(\vec{s}_i^-,\vec{s}_j^+)} +
\ds{\frac{d\Gamma}{d\hat{s}}(-\vec{s}_i^-,\vec{s}_j^+) \Bigg)} +
\ds{\Bigg( \frac{d\Gamma}{d\hat{s}}(\vec{s}_i^-,-\vec{s}_j^+)} +
\ds{\frac{d\Gamma}{d\hat{s}}(-\vec{s}_i^-,-\vec{s}_j^+)\Bigg)}}~,
\eea where $i,j=L,~N,~T$, and the first index $i$ corresponds to
lepton while the second index $j$ corresponds to antilepton,
respectively. After doing the straightforward calculation we
obtain the following expressions for $P_{ij}(\hat{s})$:
\begin{eqnarray} \label{PLLf} P_{LL} \es \frac{m_{B_{s}}^2}{3 \hat{r}_{\phi}
\hat{s} \Delta} \, \mbox{\rm Re} \bigg\{
- 24 m_{B_{s}}^2 \hat{m}_\ell^2 \hat{s} \lambda
\Big[ (B_1-D_1) (B_3^\ast - D_3^\ast) \Big] \nnb \\
\ar 12 m_{B_{s}}^3 \hat{m}_\ell \hat{s} \lambda (1-\hat{r}_{\phi})
\Big[ 2 m_{B_{s}} \hat{m}_\ell
(B_2 - D_2) (B_3^\ast - D_3^\ast) \Big] \nnb \\
\ek 8 m_{B_{s}}^4 \hat{r}_{\phi} \hat{s}^2 \lambda (1+3 v^2) (\vel
B_0 \ver^2 + \vel C_1 \ver^2) + 12 m_{B_{s}}^4 \hat{m}_\ell^2
\hat{s}^2 \lambda
\vel B_3 - D_3 \ver^2 \nnb \\
\ar 8 m_{B_{s}}^2 \hat{m}_\ell^2 \lambda (4 - 4 \hat{r}_{\phi} -
\hat{s}) (B_1 D_2^\ast + B_2 D_1^\ast)  - 32 \hat{m}_\ell^2
(\lambda + 3 \hat{r}_{\phi} \hat{s}) B_1 D_1^\ast \nnb \\
\ek 8 m_{B_{s}}^4 \hat{m}_\ell^2 \lambda [\lambda + 3
(1-\hat{r}_{\phi})^2] B_2 D_2^\ast  - 64 m_{B_{s}}^4
\hat{m}_\ell^2
\hat{r}_{\phi} \hat{s} \lambda B_0 C_1^\ast \nnb \\
\ar 8 m_{B_{s}}^2 \lambda [\hat{s} - \hat{s} (\hat{r}_{\phi} +
\hat{s}) - 3 \hat{m}_\ell^2
(2 - 2 \hat{r}_{\phi} - \hat{s})] (B_1 B_2^\ast + D_1 D_2^\ast) \nnb \\
\ek m_{B_{s}}^4 \hat{s} \lambda [\lambda (1+3 v^2) - 3
(1-\hat{r}_{\phi})^2 (1-v^2)]
(\vel B_2 \ver^2 + \vel D_2 \ver^2) \nnb \\
\ar 4 [6 \hat{m}_\ell^2 (\lambda + 6 \hat{r}_{\phi}\hat{s}) -
\hat{s} (\lambda + 12 \hat{r}_{\phi}\hat{s})] (\vel B_1 \ver^2 +
\vel D_1 \ver^2)\bigg\} ~,\\ \nnb\\
 \label{PLNf}P_{LN} \es \frac{\pi
m_{B_{s}}^2}{2\hat{r}_{\phi}\Delta} \sqrt{\frac{\lambda}{\hat{s}}}
\, \mbox{\rm Im} \bigg\{
- 4 m_{B_{s}}^4 \hat{m}_\ell \lambda (1-\hat{r}_{\phi}) B_2 D_2^\ast \nnb \\
\ar 2 m_{B_{s}}^4 \hat{m}_\ell \hat{s} \lambda B_2 B_3^\ast - 2
m_{B_{s}}^4 \hat{m}_\ell \hat{s} \lambda
\Big[ B_3 D_2^\ast + (B_2 + D_2) D_3^\ast \Big] \nnb \\
\ek 2 m_{B_{s}}^2 \hat{m}_\ell \hat{s} (1+3 \hat{r}_{\phi}
-\hat{s}) \Big( B_1 B_2^\ast - D_1 D_2^\ast \Big) - 4 \hat{m}_\ell
(1-\hat{r}_{\phi}-\hat{s})
 B_1 D_1^\ast  \nnb \\
\ek 2 m_{B_{s}}^2 \hat{m}_\ell \hat{s} (1-\hat{r}_{\phi}-\hat{s})
(B_1 + D_1) (B_3^\ast - D_3^\ast) \nnb \\
\ar 2 m_{B_{s}}^2 \hat{m}_\ell [\lambda + (1-\hat{r}_{\phi})
(1-\hat{r}_{\phi}-\hat{s})] \Big( B_2 D_1^\ast + B_1 D_2^\ast
\Big) \bigg\} ~, \\\nnb\\
 \label{PNLf}P_{NL}\es-P_{LN}~,\\ \nnb\\
\label{PLTf} P_{LT} \es \frac{\pi m_{B_{s}}^2
v}{\hat{r}_{\phi}\Delta} \sqrt{\frac{\lambda}{\hat{s}}} \,
\mbox{\rm Re} \bigg\{
m_{B_{s}}^4 \hat{m}_\ell \lambda (1-\hat{r}_{\phi})
\vel B_2 - D_2 \ver^2 \nnb \\
\ek 8 m_{B_{s}}^2 \hat{m}_\ell \hat{r}_{\phi} \hat{s} \Big( B_0
B_1^\ast - C_1 D_1^\ast \Big)+ m_{B_{s}}^4 \hat{s} \lambda \hat{m}_\ell B_2 B_3^\ast  \nnb \\
\ek m_{B_{s}}^4 \hat{m}_\ell \hat{s} \lambda \Big(B_2 D_3^\ast +
B_3 D_2^\ast - D_2 D_3^\ast \Big)  + \hat{m}_\ell
(1-\hat{r}_{\phi} -\hat{s})
\vel B_1 - D_1 \ver^2 \nnb \\
\ar m_{B_{s}} \hat{s} (1-\hat{r}_{\phi} -\hat{s}) \Big[
- m_{B_{s}} \hat{m}_\ell (B_1 - D_1) (B_3^\ast - D_3^\ast) \Big] \nnb \\
\ek m_{B_{s}}^2 \hat{m}_\ell [\lambda + (1-\hat{r}_{\phi})
(1-\hat{r}_{\phi} -\hat{s})] (B_1 - D_1) (B_2^\ast - D_2^\ast)
\bigg\}~, \\\nnb\\
 \label{PTLf}
 P_{TL} \es \frac{\pi m_{B_{s}}^2
v}{\hat{r}_{\phi}\Delta} \sqrt{\frac{\lambda}{\hat{s}}} \,
\mbox{\rm Re} \bigg\{
m_{B_{s}}^4 \hat{m}_\ell \lambda (1-\hat{r}_{\phi})
\vel B_2 - D_2 \ver^2 \nnb \\
\ar 8 m_{B_{s}}^2 \hat{m}_\ell \hat{r}_{\phi} \hat{s} \Big( B_0
B_1^\ast - C_1 D_1^\ast \Big)+ m_{B_{s}}^4 \hat{s} \lambda \hat{m}_\ell B_2 B_3^\ast  \nnb \\
\ek m_{B_{s}}^4 \hat{m}_\ell \hat{s} \lambda \Big(B_2 D_3^\ast +
B_3 D_2^\ast - D_2 D_3^\ast \Big)  + \hat{m}_\ell
(1-\hat{r}_{\phi} -\hat{s})
\vel B_1 - D_1 \ver^2 \nnb \\
\ek m_{B_{s}} \hat{s} (1-\hat{r}_{\phi} -\hat{s}) \Big[
 m_{B_{s}} \hat{m}_\ell (B_1 - D_1) (B_3^\ast - D_3^\ast) \Big] \nnb \\
\ek m_{B_{s}}^2 \hat{m}_\ell [\lambda + (1-\hat{r}_{\phi})
(1-\hat{r}_{\phi} -\hat{s})] (B_1 - D_1) (B_2^\ast - D_2^\ast)
\bigg\}~, \\\nnb\\
 \label{PNTf}
 P_{NT} \es \frac{2 m_{B_{s}}^2 v}{3
\hat{r}_{\phi}\Delta} \, \mbox{\rm Im} \bigg\{
4 \lambda \Big(B_1 D_1^\ast + m_{B_{s}}^4 \lambda B_2 D_2^\ast
\Big) -16 m_{B_{s}}^4 \hat{s} \lambda \hat{r}_{\phi} B_0
C_1^\ast\nnb \\\ek 4 m_{B_{s}}^2 \lambda
(1-\hat{r}_{\phi}-\hat{s}) \Big(B_1 D_2^\ast + B_2 D_1^\ast \Big)
\bigg\}~,\\ \nnb \\
 \label{PTNf}P_{TN}\es-P_{NT}~,\\\nnb \\
\label{PNNf}
 P_{NN} \es \frac{2 m_{B_{s}}^2}{3
\hat{r}_{\phi}\Delta} \, \mbox{\rm Re} \bigg\{
- 24 \hat{m}_\ell^2 \hat{r}_{\phi} (\vel B_1 \ver^2 + \vel D_1
\ver^2)+16 m_{B_{s}}^4 \hat{s} \lambda
 \hat{r}_{\phi} v^2  B_0 C_1^\ast \nnb \\
\ar 6 m_{B_{s}}^2 \hat{m}_\ell^2 \lambda \Big[- 2 B_1 (B_2^\ast +
B_3^\ast - D_3^\ast) +
2 D_1 (B_3^\ast - D_2^\ast - D_3^\ast) \Big] \nnb \\
\ar 6 m_{B_{s}}^3 \hat{m}_\ell \lambda (1-\hat{r}_{\phi}) \Big[ 2
m_{B_{s}} \hat{m}_\ell (B_2 - D_2) (B_3^\ast - D_3^\ast) \Big]
  \nnb \\
\ar 6 m_{B_{s}}^4 \hat{m}_\ell^2 \lambda (2+2
\hat{r}_{\phi}-\hat{s}) (\vel B_2 \ver^2 + \vel D_2 \ver^2)  + 6
m_{B_{s}}^4 \hat{m}_\ell^2 \hat{s} \lambda
\vel B_3 - D_3 \ver^2 \nnb \\
\ar m_{B_{s}}^2 \lambda [3 (2 - 2 \hat{r}_{\phi} - \hat{s}) - v^2
(2 - 2 \hat{r}_{\phi} + \hat{s})]
(B_1 D_2^\ast + B_2 D_1^\ast) \nnb \\
\ek m_{B_{s}}^4 \lambda \Big[ (3+v^2) \lambda + 3 (1-v^2)
(1-\hat{r}_{\phi})^2 \Big]
B_2 D_2^\ast \nnb \\
\ek 2 [6 \hat{r}_{\phi} \hat{s} (1-v^2) + \lambda (3-v^2)] B_1
D_1^\ast \bigg\}~,\\\nnb\\
 \label{PTTf}
 P_{TT} \es \frac{2 m_{B_{s}}^2}{3 \hat{r}_{\phi}
\hat{s}\Delta} \, \mbox{\rm Re} \bigg\{
8 m_{B_{s}}^4 \hat{r}_{\phi} \hat{s} \lambda \Big[ 4
\hat{m}_\ell^2 (\vel B_0 \ver^2 + \vel C_1 \ver^2)
+ 2 \hat{s} B_0 C_1^\ast \Big] \nnb \\
\ek 6 m_{B_{s}}^2 \hat{m}_\ell^2 \hat{s} \lambda
\Big[ -2 (B_1 - D_1) (B_3^\ast - D_3^\ast) \Big] \nnb \\
\ek 6 m_{B_{s}}^3 \hat{m}_\ell \hat{s} \lambda (1-\hat{r}_{\phi})
\Big[  2 m_{B_{s}} \hat{m}_\ell
(B_2 - D_2) (B_3^\ast - D_3^\ast) \Big] \nnb \\
\ek 6 m_{B_{s}}^4 \hat{m}_\ell^2 \hat{s}^2 \lambda \vel B_3 - D_3
\ver^2  + 4 m_{B_{s}}^2 \hat{m}_\ell^2 \lambda
(4 - 4 \hat{r}_{\phi} - \hat{s}) (B_1 B_2^\ast + D_1 D_2^\ast) \nnb \\
\ar 2 \hat{s} [6 \hat{r}_{\phi} \hat{s} (1-v^2) + \lambda (1-3
v^2)]
B_1 D_1^\ast \nnb \\
\ek 2 m_{B_{s}}^4 \hat{m}_\ell^2 \lambda [\lambda + 3
(1-\hat{r}_{\phi})^2]
(\vel B_2 \ver^2 + \vel D_2 \ver^2) \nnb \\
\ek m_{B_{s}}^2 \hat{s} \lambda [2 - 2 \hat{r}_{\phi} + \hat{s} -
3 v^2 (2 - 2 \hat{r}_{\phi} - \hat{s})]
(B_1 D_2^\ast + B_2 D_1^\ast) \nnb \\
\ek 8 \hat{m}_\ell^2 (\lambda - 3 \hat{r}_{\phi}\hat{s})
(\vel B_1 \ver^2 + \vel D_1 \ver^2) \nnb \\
\ek m_{B_{s}}^4 \hat{s} \lambda \Big[ (1+3 v^2) \lambda - 3
(1-v^2) (1-\hat{r}_{\phi})^2 \Big] B_2 D_2^\ast
 \bigg\}~. \eea
\par The analytical dependence of the double--lepton polarizations
on the fourth quark mass($m_{t'}$) and the product of quark mixing
matrix elements ($V_{t^\prime b}^\ast V_{t^\prime
s}=r_{sb}e^{i\phi_{sb}}$) are studied in the next section.

\section{Effects of the  fourth-generation }

 As we mentioned in the introduction, the inclusion of
the  fourth-generation in  the Standard Model (SM4)  does not lead
to  new operators  in the ${\cal H}_{\rm eff}$ and  all Wilson
coefficients  receive additional terms as
$\frac{\lambda_{t'}}{\lambda_{t}} C^{\rm SM4}_i$ either via
virtual exchange of the fourth-generation up-type quark $t'$
$(C_3,...,C_{10})$ or via using the unitarity of CKM matrix
$(C_1,C_2)$ . Consequently, one can write the new effective
Hamiltonian as:
\begin{equation}
{\cal H}_{\rm eff} = -\frac{ G_F}{\sqrt{2}} V_{tb}
V^*_{ts}\sum_{i=1}^{10} C^{\rm new}_i(\mu) {\cal O}_i(\mu),
\end{equation}
 where $C^{new}_i$ are:
\begin{eqnarray}\label{Wilson}
C^{\rm new}_i (\mu)  &=& C_i(\mu) + \frac{\lambda_{t'}}{\lambda_{t}}
C^{\rm SM4}_i(\mu),~~~~~~~~~~~~i=1\ldots10.
\end{eqnarray}
In the above equation, $\lambda_f=V_{f b}^\ast V_{fs}$ and
$\lambda_{t'}$ can be parameterized as:
\begin{equation}
\lambda_{t'}=V_{t'b}V_{t's}^{*}=r_{sb}e^{i\phi_{sb}}.
\end{equation}

 Now by using the above effective
Hamiltonian, we can reobtain the one-loop matrix elements of $b
\rightarrow s \ell^+ \ell^-$ by replacing $C^{\rm
eff}_i(\tilde{C}^{\rm eff}_i)$ with ${C^{\rm eff\,
new}_i}({{\tilde{C}}^{\rm eff\, \rm new}_i})$ in
Eq.(\ref{bdecay}), where  ${C^{\rm eff\, new}_i}$ and
${{\tilde{C}}^{\rm eff\, \rm new}_i}$are given as:
\begin{eqnarray}
C^{\rm eff\,\rm new}_i (\mu)  &=& C^{\rm eff}_i(\mu) +
\frac{\lambda_{t'}}{\lambda_{t}} C^{\rm eff\,\,\rm
SM4}_i(\mu),~~~~~~~~~~~~i=7,\nnb \\ \tilde{C}^{\rm eff\,\rm new}_i
(\mu) &=& \tilde{C}^{\rm eff}_i(\mu) +
\frac{\lambda_{t'}}{\lambda_{t}} \tilde{C}^{\rm eff\,\,\rm
SM4}_i(\mu),~~~~~~~~~~~~i=9,10.
\end{eqnarray}
Here the effective Wilson coefficients ${C}^{\rm eff\,\,\rm
SM4}_i$ and $\tilde{C}^{\rm eff\,\,\rm SM4}_i$ are defined in the
same way as Eqs.(\ref{c9}) by substituting $C_i$ with $C_i^{SM4}$.
It is worth nothing that the explicit forms of ${C}^{\rm
eff\,\,\rm SM4}_i$ and $\tilde{C}^{\rm eff\,\,\rm SM4}_i$ can also
be found   from the corresponding Wilson coefficients in SM by
replacing $m_t\rightarrow m_{t'}$ \cite{R23}. Based on the
preceding explanations, in order to obtain the matrix element and
the double-lepton polarization  asymmetries for $B_s \rightarrow
\phi \ell^+ \ell^-$ decay in the presence of the
fourth-generation, one should replace $C^{\rm
eff}_i(\tilde{C}^{\rm eff}_i)$ with ${C^{\rm eff\,
new}_i}({{\tilde{C}}^{\rm eff\, \rm new}_i})$ in  all equations of
the previous section.

The unitary quark mixing matrix is now $4\times4$ which can be
written in terms of $6$ mixing angles and $3$ CP violating phases.
The relevant  elements of this matrix for $b \rightarrow s$
transition satisfy the relation:
 \begin{equation}\label{unitary}
\lambda_{u}+\lambda_{c}+\lambda_{t}+\lambda_{t'}=0.
\end{equation}
Consequently, as required by GIM mechanism, the factor
$\lambda_{t} C_{i}^{\rm new}$ should be modified to  $\lambda_{t}
C_{i}$ when $ m_{t'}\rightarrow m_t $ or $\lambda_{t'}\rightarrow
0$. We can easily check the validity of this condition by using
Eq.(\ref{unitary}):
\begin{eqnarray}
\lambda_{t} C_{i}^{\rm new}=\lambda_{t}  C_{i}+\lambda_{t'}
C_{i}^{\rm SM4}&=&-(\lambda_{u}+\lambda_{c}) C_{i}+\lambda_{t'}(
C_{i}^{\rm SM4}-C_{i}) \nonumber \\
 &=& -(\lambda_{u}+\lambda_{c}) C_{i}\nonumber \\
 &=& \lambda_{t} C_{i}.
\end{eqnarray}

The numerical analysis of the dependence of the double--lepton
polarizations on the fourth quark mass ($m_{t'}$) and the product
of quark mixing matrix elements ($V_{t^\prime b}^\ast V_{t^\prime
s}=r_{sb}e^{i\phi_{sb}}$) are presented in the next section.
\section{Results and Discussions}\label{results}

The main input parameters in the calculations are the form factors
for which  we have chosen the predictions of light cone QCD sum
rule method  \cite{R31,R32},   as pointed out in
 section II.
 Besides the form factors, we  use the other input parameters as
follow: \bea && m_{B_s} =5.37 \, \mbox{GeV} \, , \, m_b =4.8 \,
\mbox{GeV} \,, \,m_c=1.5 \, \mbox{GeV} \, , \, m_{\tau} =1.77 \,
\mbox{GeV} \,,\,
 \nnb \\ &&m_{\mu} =0.105 \, \mbox{GeV},\, m_{\phi}=1.020 \,
  \mbox{GeV} \,\, ,\,|V_{tb}V_{ts}^*|=0.0385\,,\,\alpha^{-1}=129\,,\,\nnb\\
&&G_f=1.166\times10^{-5}\, {\mbox{GeV}}^{-2}\,,\,\tau_{B_s}=1.46
\times 10^{-12}\,s\,. \eea

  In order to present a quantitative analysis of the double-lepton polarization asymmetries, the values of fourth-generation
  parameters are needed. Considering the
 experimental values of $B\longrightarrow X_s \gamma$ and $B\longrightarrow X_s \ell^+\ell^-$
   decays the value
 of the  $r_{sb}$ parameter is restricted to the  range $\{.01-.03\}$
 for $\phi_{sb}\sim\{0^\circ-360^\circ\}$ and $m_{t'}\sim\{200-600\}$
 GeV\cite{Arhrib:2002md,Zolfagharpour:2007ez2}.
  Using the  $B_s$ mixing parameter $\Delta m_{B_s}$, a sharp restriction on $\phi_{sb}$
   has been obtained ($\phi_{sb}\sim 90^\circ$)\cite{Hou:2006jy}. Therefore in our following numerical analysis,
 the corresponding values of above ranges are:
$r_{sb}=\{.01,~.02,~.03\},\phi_{sb}=\{60^\circ,~ 90^\circ,
~120^\circ\},m_{t'}=175\leq m_{t'} \leq600$.

It is clear from the expressions of all nine double--lepton
polarization asymmetries that they depend on the momentum transfer
$q^2$ and the new parameters $(m_{t'}$, $r_{sb}$, $\phi_{sb})$.
Consequently, it may  be experimentally  difficult to investigate
these dependencies at the same time. One way to deal with this
problem is to integrate over $q^2$ and study the averaged
double-lepton polarization asymmetries. The average of $P_{ij}$
over $q^2$ is defined as:

 \bea \la P_{ij} \ra = \frac{\ds \int_{4
\hat{m}_\ell^2}^{(1-\sqrt{\hat{r}_{\phi}})^2} P_{ij} \frac{d{\cal
B}}{d \hat{s}} d \hat{s}} {\ds \int_{4
\hat{m}_\ell^2}^{(1-\sqrt{\hat{r}_{\phi}})^2} \frac{d{\cal B}}{d
\hat{s}} d \hat{s}}~. \eea

We have used the above formula and depicted the dependency of $\la
P_{ij}\ra$ on the fourth-generation parameters in
Fig.[\ref{PLL}-\ref{PTT}]. In the following, we compare our
results for $B_s\rightarrow \phi \ell^+\ell^-$ decay with the
results of Ref.\cite{Bashiry:2007tf} for $B\rightarrow K
\ell^+\ell^-$ decay. Since the overall  behavior of $\la
P_{ij}\ra$  versus $m_{t'},r_{sb}$ and $\phi_{sb}$ are almost the
same as that of $B\rightarrow K \ell^+\ell^-$ decay, we discuss
the differences of these two decays and some aspects which have
not been discussed in Ref.\cite{Bashiry:2007tf}:

\begin{itemize}
\item \textbf{Figrue(\ref{PLL})}: Similar to the $B \rightarrow K \mu^+\mu^-$   decay,
  $\la P_{LL}\ra$ is not sensitive to the
fourth-generation quark parameters, therefore the  $\la P_{LL}\ra$
plots for $\mu$ channel have been omitted. However, for the $\tau$
channel, the maximum deviation from SM is about $50 \%$
  which can be seen at $m_{t'}\sim 600 GeV$. In comparison with  the
  results   of Ref.\cite{Bashiry:2007tf},  it is understood that the  deviation from SM for
  $B_s \rightarrow \phi \tau^+\tau^-$  is twice that of $B \rightarrow K \tau^+\tau^-$
  decay. Therefore, the magnitude of $\la P_{LL}\ra$ in $B_s \rightarrow \phi\tau^+\tau^-$ compared with that in $B \rightarrow K\tau^+\tau^-$decay  has more chance to show the existence of the fourth-generation.

\item \textbf{Figrue(\ref{PLN})}:
The value of  $\la P_{LN}\ra_{max}$ for $\mu$  channel is about
0.04 which is four times greater than that for $B \rightarrow K$
decay. However, for $\tau$ channel such value is at most around
$0.3 $ which is approximately equal to the maximum value of $\la
P_{LN}\ra$ for $B \rightarrow K $ decay. Furthermore, in $\mu$ and
$\tau$ channels by increasing $r_{sb}$ and keeping the values of
$\phi_{sb}$ fixed, the maximum deviation from SM occurs at smaller
values of $m_{t'}$. This result can be interesting since the
maximum deviation from SM happens for $r_{sb}\sim \{0.02 -0.03\}$
and $m_{t'}\sim\{300-400\}$GeV. Therefore, the  new generation has
a chance to be observed around  $m_{t'}\sim\{300-400\}$GeV. Our
analysis shows that to measure the effect of the
fourth-generation in $\la P_{LN}\ra$, the $\tau$ channel of $B_s
\rightarrow \phi $ and $B \rightarrow K$ are more important than
$\mu$ channel of these decays, knowing that in the $\mu $ channel
the $B_s \rightarrow \phi $ decay is more significant than the $B
\rightarrow K $ decay.

\item \textbf{Figrue(\ref{PLT})}:  For $\mu$ channel, the magnitude of $\la
P_{LT}\ra$ in  $B_s \rightarrow \phi$ decay changes at most about
$80\%$ compared with the SM prediction, while the maximum change
in $B \rightarrow K$ decay reaches up to  $60\%$. For $\tau$ case,
unlike $B \rightarrow K$ decay, the magnitude of $\la P_{LT}\ra$
in $B_s \rightarrow \phi$ transition exhibits the strong
dependence on the fourth quark mass $(m_{t'})$ and the product of
quark mixing matrix elements $(|V_{t'b}V_{t's}^*|=r_{sb})$. As
seen from Fig.(\ref{PLT}) the maximum deviation from SM in $\tau$
channel is much more than that in $\mu$ channel. Therefore for
establishing the fourth generation of quarks the measurement of
$\la P_{LT}\ra$ for $B_s \rightarrow \phi\tau^{+}\tau^{-}$  decay
is more suitable than such measurement for $B_s \rightarrow
\phi\mu^{+}\mu^{-}$ and $B \rightarrow K \mu^{+}\mu^{-}$  decays .

\item \textbf{Figrue(\ref{PTL})}: It is seen from Eqs.(\ref{PLTf}) and (\ref{PTLf}) that
 contrary to $B \rightarrow K$ decay, $P_{TL}$   is
neither symmetric nor anti-symmetric under the exchange of
subscripts L and T which leads to different values for $P_{TL}$
and $P_{LT}$.
 For
$\mu$ channel, the magnitude of $\la P_{TL}\ra$ in $B_s
\rightarrow \phi$ decay changes at most about $40\%$ compared with
the SM prediction, while the maximum change in the case of  $B
\rightarrow K$ decay reaches up to  $60\%$.
 For $\tau$ case, unlike $B \rightarrow K$ decay, the
magnitude of $\la P_{TL}\ra$ in $B_s \rightarrow \phi$ transition
changes at most about $60\%$ compared with the SM prediction.
Therefore, in the measurement  of $\la P_{TL}\ra$, the  decays
 $B_s\rightarrow\phi \ell^{+}\ell^{-}$($\ell=\mu,\tau$) and $B
\rightarrow K \mu^{+}\mu^{-}$ have the same significance  for
finding the new generation of quarks.

\item \textbf{Figrue(\ref{PTN})}: By comparing this figure with Fig.(\ref{PLN}), one can find out that
 the overall behavior of $\la P_{TN}\ra$
 and $\la P_{LN}\ra$ are the same.
 Furthermore, the magnitude of   $\la P_{TN}\ra_{max}$ for
$\mu$  channel is about 0.22 which is four times smaller than that
for $B \rightarrow K$ decay and for $\tau$ channel such value is
at most around $0.0075 $ which is approximately ten times smaller
than  $\la P_{TN}\ra_{max}$ for $B \rightarrow K $ decay. Although
the measurement of $\la P_{TN}\ra$ in $B \rightarrow K
\tau^+\tau^-$ decay for finding the new generation is useful,
such  measurement in the decays $B_s \rightarrow \phi \mu^+\mu^-$
and $B \rightarrow K \mu^+\mu^-$ are more significant.

\item \textbf{Figrue(\ref{PNN})}: For both  $\mu$ and $\tau$
channels in $B_s \rightarrow \phi$ decay, the values of $\la
P_{NN}\ra$ show  stronger dependence on the fourth-generation
parameters $(m_{t'},r_{sb},\phi_{sb})$ in comparison with those in
$B \rightarrow K$ decay. Furthermore,  the situation for $\tau$
channel is even more interesting than $\mu$ channel, since for
fixed values of $\phi_{sb}$ and $r_{sb}$, an increase in  $m_{t'}$
changes the sign of $\la P_{NN}\ra$. So, for $B_s \rightarrow
\phi$ decay, the study of the magnitude and the sign of $\la
P_{NN}\ra$ for $\tau$ channel and the magnitude of this asymmetry
in $\mu$ channel can serve as good tests for discovering the new
physics beyond the SM. It should also be mentioned that for both
$\mu$ and $\tau$ channels of $B \rightarrow K$ decay in general,
 and specially for the $\mu$ channel, the deviation of $\la
P_{NN}\ra$ from SM can be a measurable quantity, even though it is
less sensitive to the fourth generation of quarks compered with
that of $B\rightarrow \phi$ decay(see Ref.\cite{Bashiry:2007tf}).

\item \textbf{Figrue(\ref{PTT})}: A comparison between this figure and an analogous figure for $B\rightarrow K \ell^+ \ell^-$
shows that the values of $\la P_{TT}\ra$  for  both $\mu$ and
$\tau$ channels in $B_s\rightarrow \phi$ decay have considerable
dependency  on the fourth-generation parameters
$(m_{t'},r_{sb},\phi_{sb})$. Therefore, compared with
 $B\rightarrow K \ell^+ \ell^-$ decay in Ref.\cite{Bashiry:2007tf}, the study of  the magnitude of $\la
P_{TT}\ra$ in
 $B_s\rightarrow \phi \ell^+ \ell^-$ provides a better opportunity to see
 the effect of
 the new physics beyond the  SM.

\end{itemize}

Finally, let us briefly discuss whether it is possible to measure
 the lepton polarization asymmetries in experiments or
 not. Experimentally, to measure an asymmetry $\lla P_{ij}\rra$ of the decay with
 branching ratio $\cal{B}$ at $n\sigma$ level, the required number
 of events (i.e., the number of $B\bar{B}$) is given by the
 formula \bea N = \frac{n^2}{{\cal
B} s_1 s_2 \la P_{ij} \ra^2}~,\nnb \eea  where $s_1$ and $s_2$ are
the efficiencies of the leptons. Typical values of the
efficiencies of the $\tau$--leptons vary from $50\%$ to $90\%$ for
their different decay modes\cite{R6016} and the error in
$\tau$--lepton polarization is estimated to be about $(10 - 15)\%$
\cite{R6017}. So, the error in measurement of the $\tau$--lepton
asymmetries is approximately $(20 - 30)\%$, and the error in
obtaining the number of events is about $50\%$.

Looking at the expression of  $N$, it can be understood that in
order to detect the lepton polarization asymmetries in the $\mu$
and $\tau$ channels at $3\sigma$ level, the minimum number of
required events are (for the efficiency of $\tau$--lepton we take
$0.5$):
\begin{itemize}
\item for $B_s \rar \phi \mu^+ \mu^-$ decay \bea N \sim \left\{
\begin{array}{ll}
10^{6}  & (\mbox{\rm for} \lla P_{LL} \rra)~,\\
10^{7}  & (\mbox{\rm for} \lla P_{NT} \rra, \lla P_{TN} \rra)~,\\
10^{8}  & (\mbox{\rm for} \lla P_{LT} \rra, \lla P_{TL} \rra, \lla P_{NN} \rra,\lla P_{TT} \rra )~,\\
 10^{9}  & (\mbox{\rm for} \lla P_{LN} \rra, \lla P_{NL} \rra)~,\\
 \end{array} \right. \nnb \eea

\item for $B_s \rar \phi \tau^+ \tau^-$ decay \bea N \sim \left\{
\begin{array}{ll}
10^{8}  & (\mbox{\rm for} \lla P_{LT} \rra, \lla P_{TL} \rra, \lla P_{NN} \rra,\lla P_{TT} \rra )~,\\
10^{9}  & (\mbox{\rm for} \lla P_{LL} \rra, \lla P_{LN} \rra, \lla P_{NL} \rra)~,\\
10^{12}  & (\mbox{\rm for} \lla P_{NT} \rra, \lla P_{TN} \rra)~.\\
\end{array} \right. \nnb \eea
\end{itemize}

Considering the above values for N and the number of $B \bar{B}$
pairs which will be produced at  LHC($\sim 10^{12}$), one can
conclude that except $\lla P_{NT} \rra$ and $\lla P_{TN} \rra$ for
$\tau$ channel, all double-lepton polarizations can be detected at
the LHC.

In summary, in this paper we have presented the analyses of the
double-lepton polarization asymmetries in $B_s \rar \phi \ell^+
\ell^-$ decay using the SM with the fourth  generation of quarks.
We found out that these asymmetries have strong dependency on the
fourth-generation parameters which can be detected at the LHC. We
compared $B_s \rar \phi \ell^+ \ell^-$decay  with  $B \rar K
\ell^+ \ell^-$ decay, and  showed that the double-lepton
polarizations of $B_s \rar \phi \ell^+ \ell^-$ are more sensitive
to the fourth-generation parameters and therefore by looking at
$B_s \rar \phi \ell^+ \ell^-$ decay, one has more chance to
investigate the correctness of the fourth generation of quarks
hypothesis in the near future.

\section{Acknowledgment}
The authors would like to thank V. Bashiry for his useful
discussions. Support of Research  Council of Shiraz University is
gratefully acknowledged.

\newpage

\begin{figure}
  \centering
  \setlength{\fboxrule}{2pt}
 \fbox{ \begin{minipage}{6 in}
        \centering
             \includegraphics[height=2.1in]{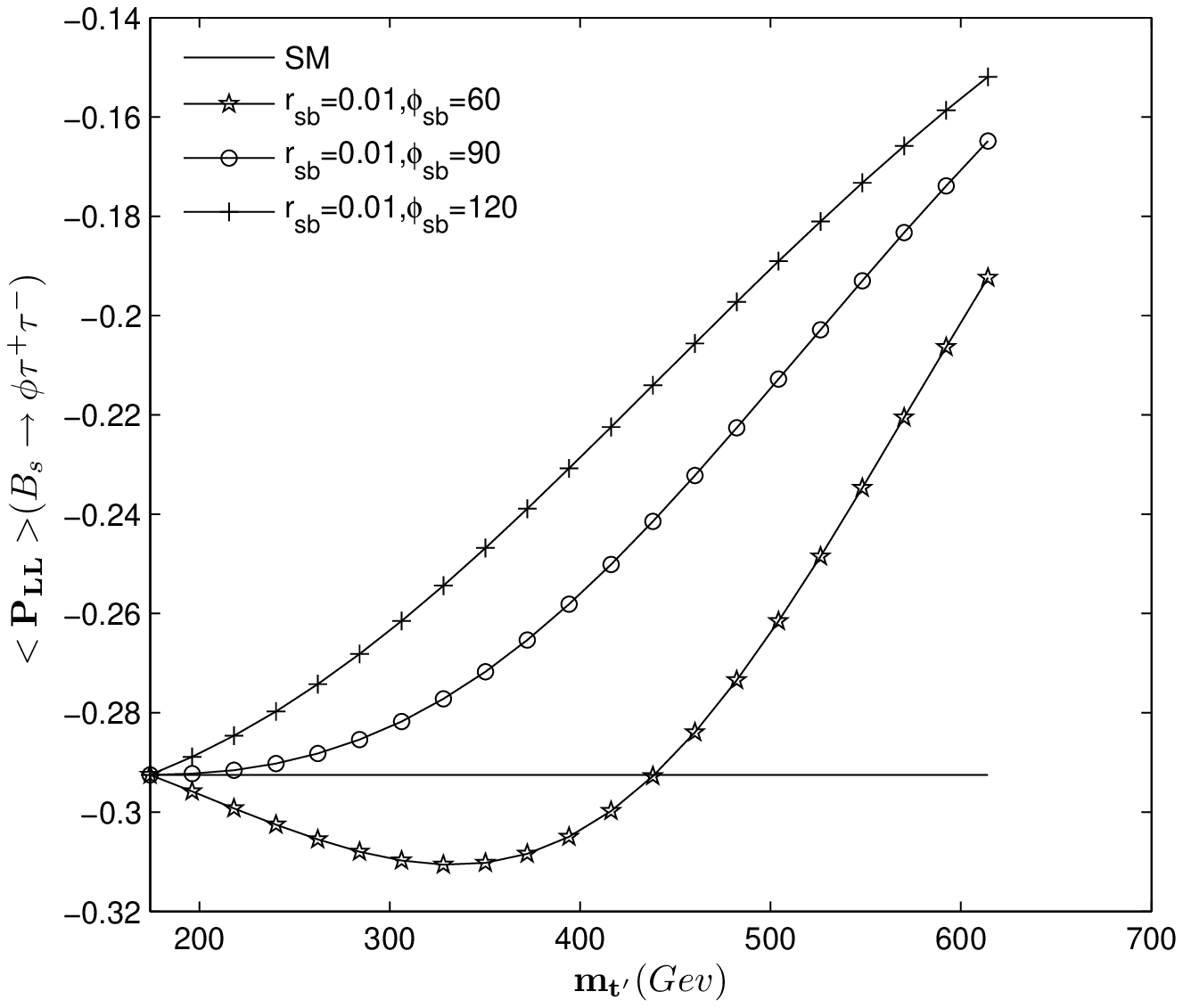}
        \includegraphics[height=2.1in]{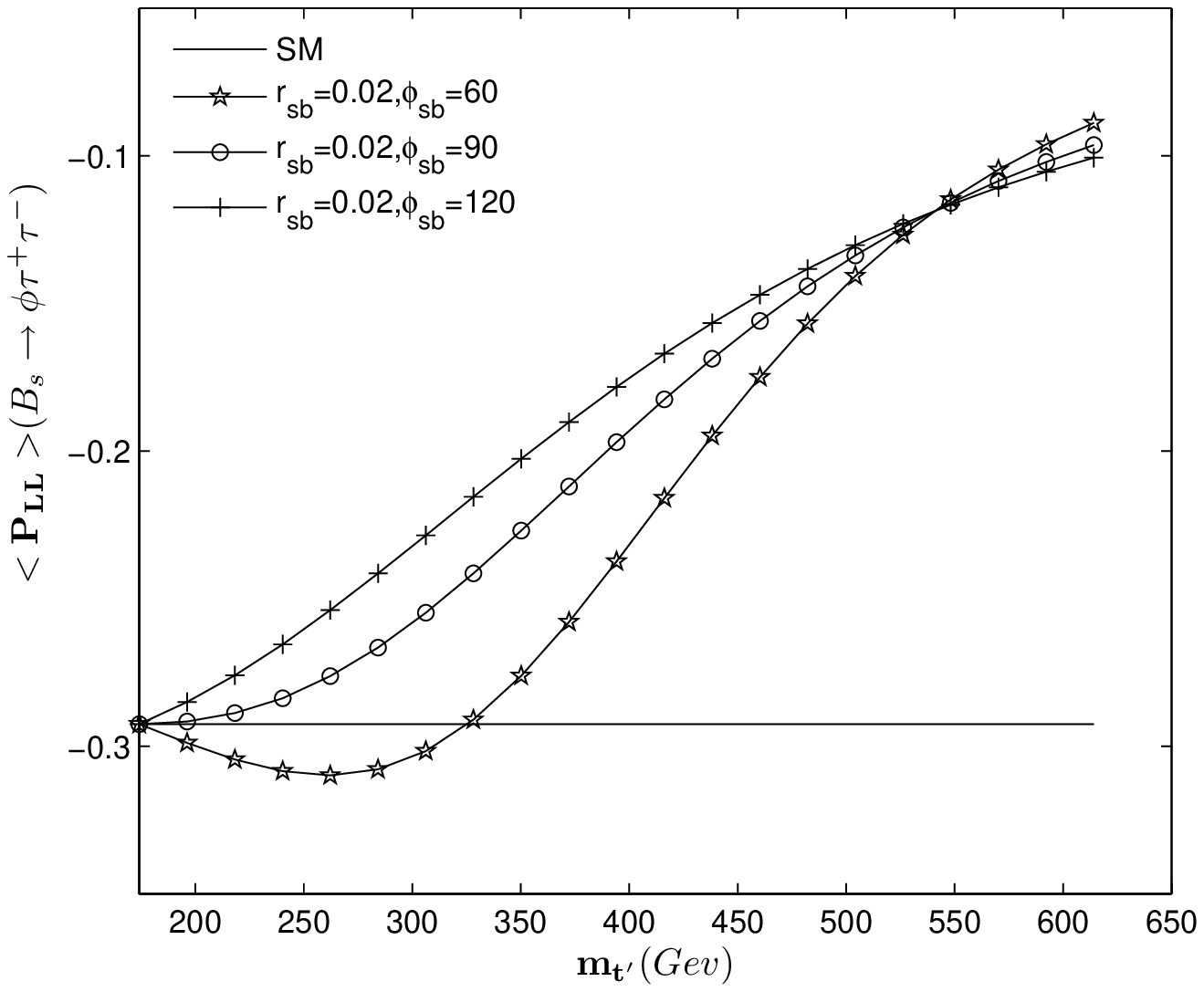}
         \includegraphics[height=2.1in]{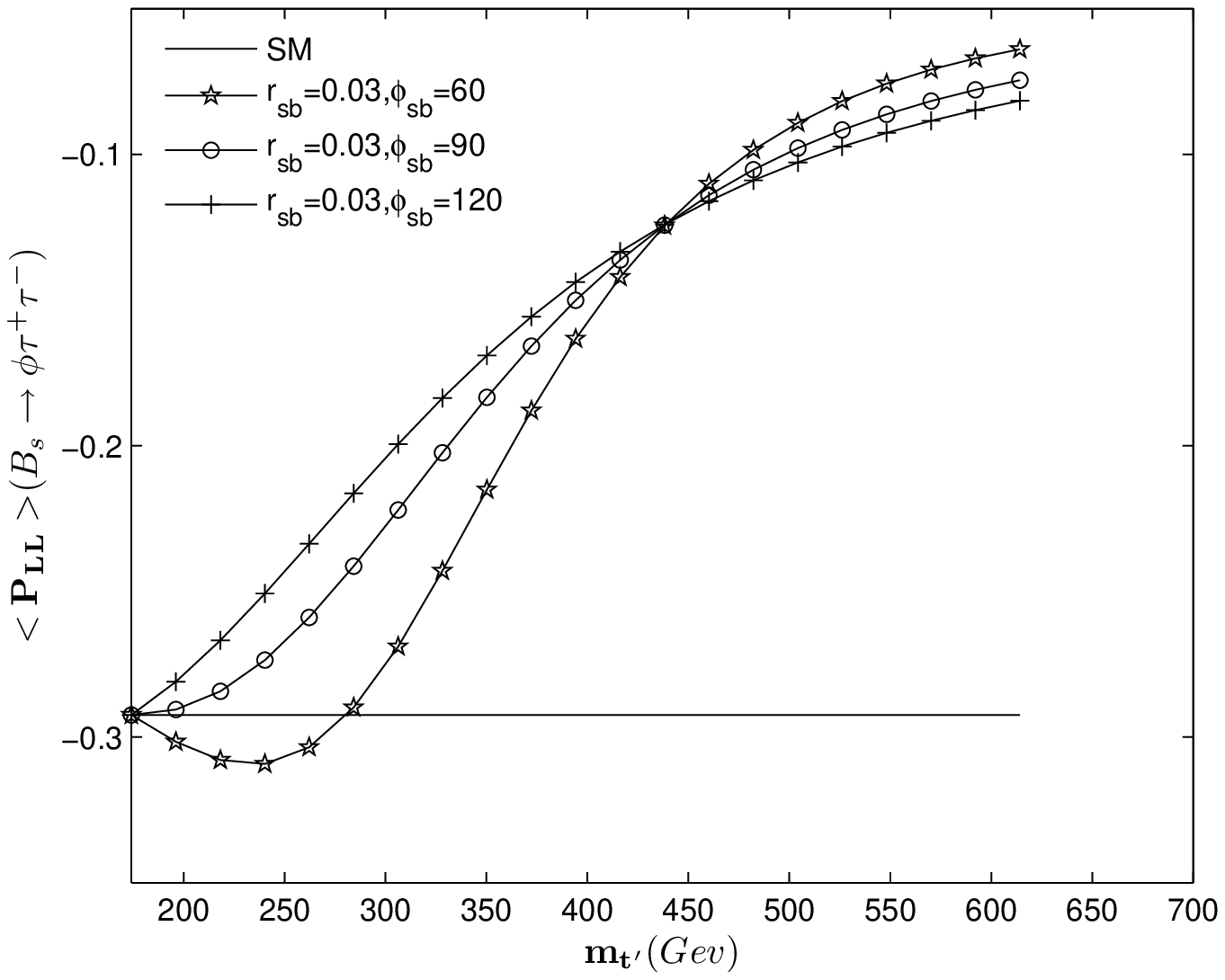}
               \caption{The dependence of the $\lla P_{LL}\rra$  on the fourth-generation quark mass
$m_{t'}$ for three different values of
 $\phi_{sb}=\{60^\circ,~ 90^\circ, ~120^\circ\}$ and $r_{sb}=\{0.01,~0.02,~0.03\}$ for the $\tau$ channel.}
\label{PLL}
           \end{minipage} }
    \end{figure}
\begin{figure}
 \centering
  \setlength{\fboxrule}{2pt}
 \fbox{ \begin{minipage}{6 in}
        \centering
        \includegraphics[height=2.1in]{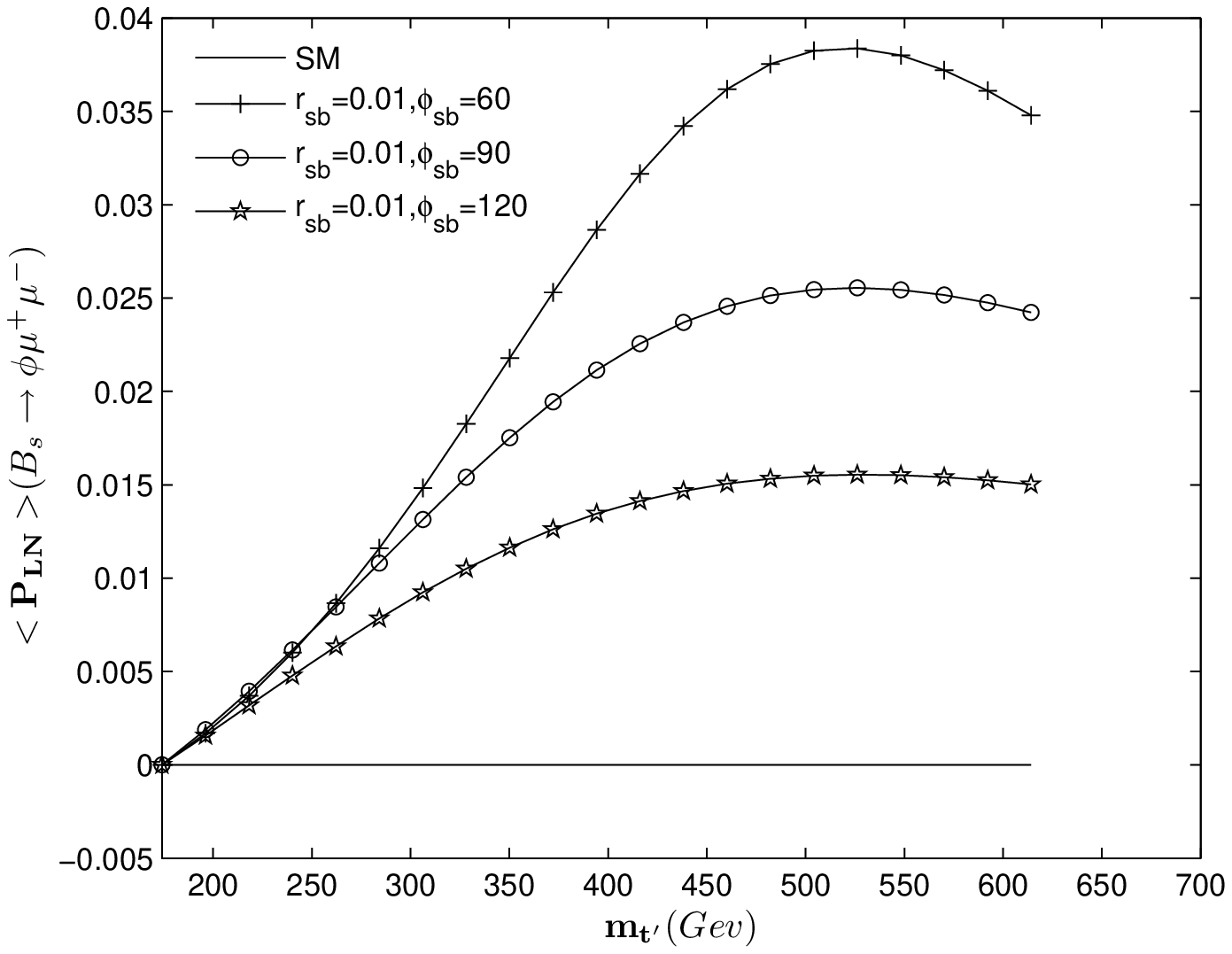}
        \includegraphics[height=2.1in]{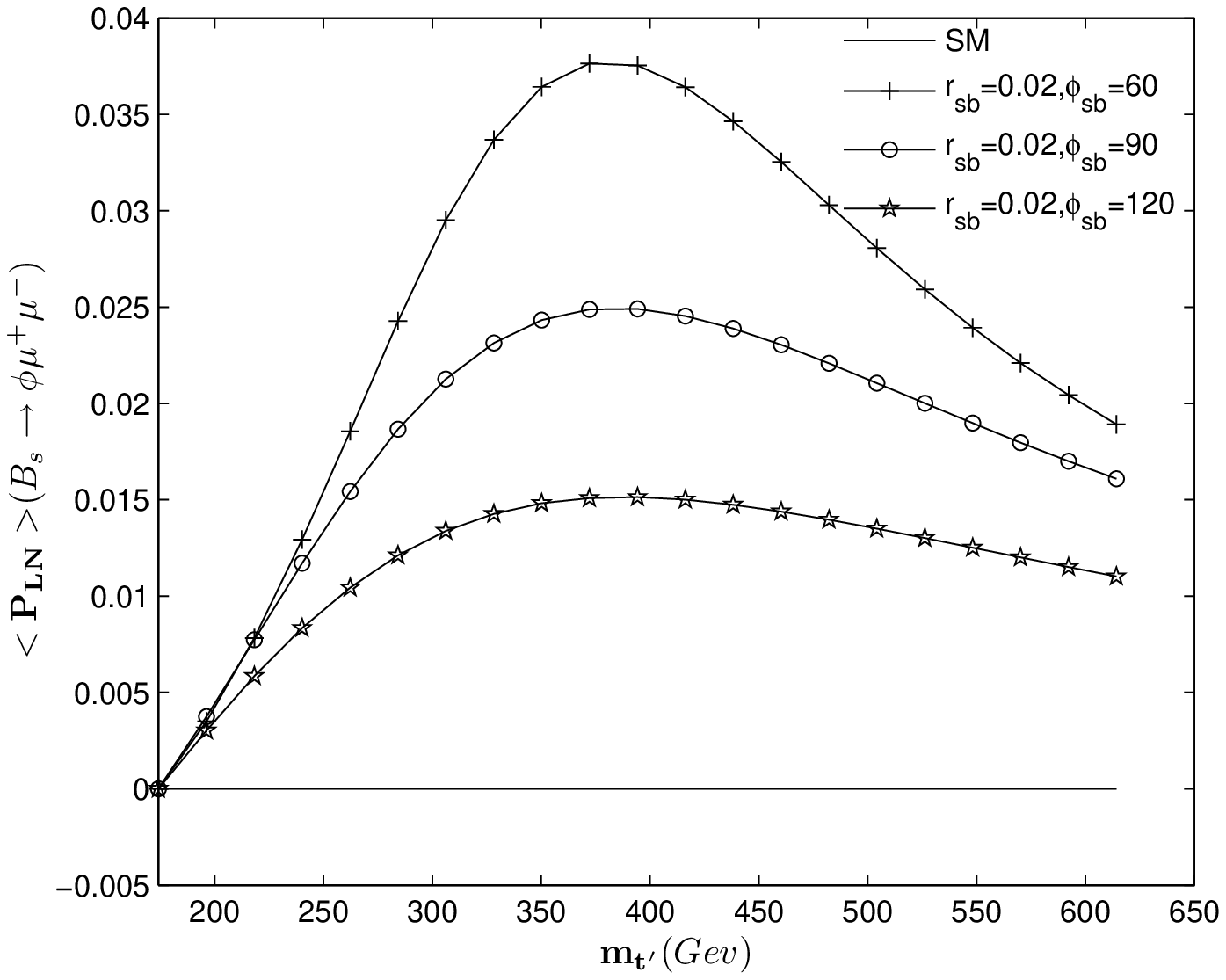}
         \includegraphics[height=2.1in]{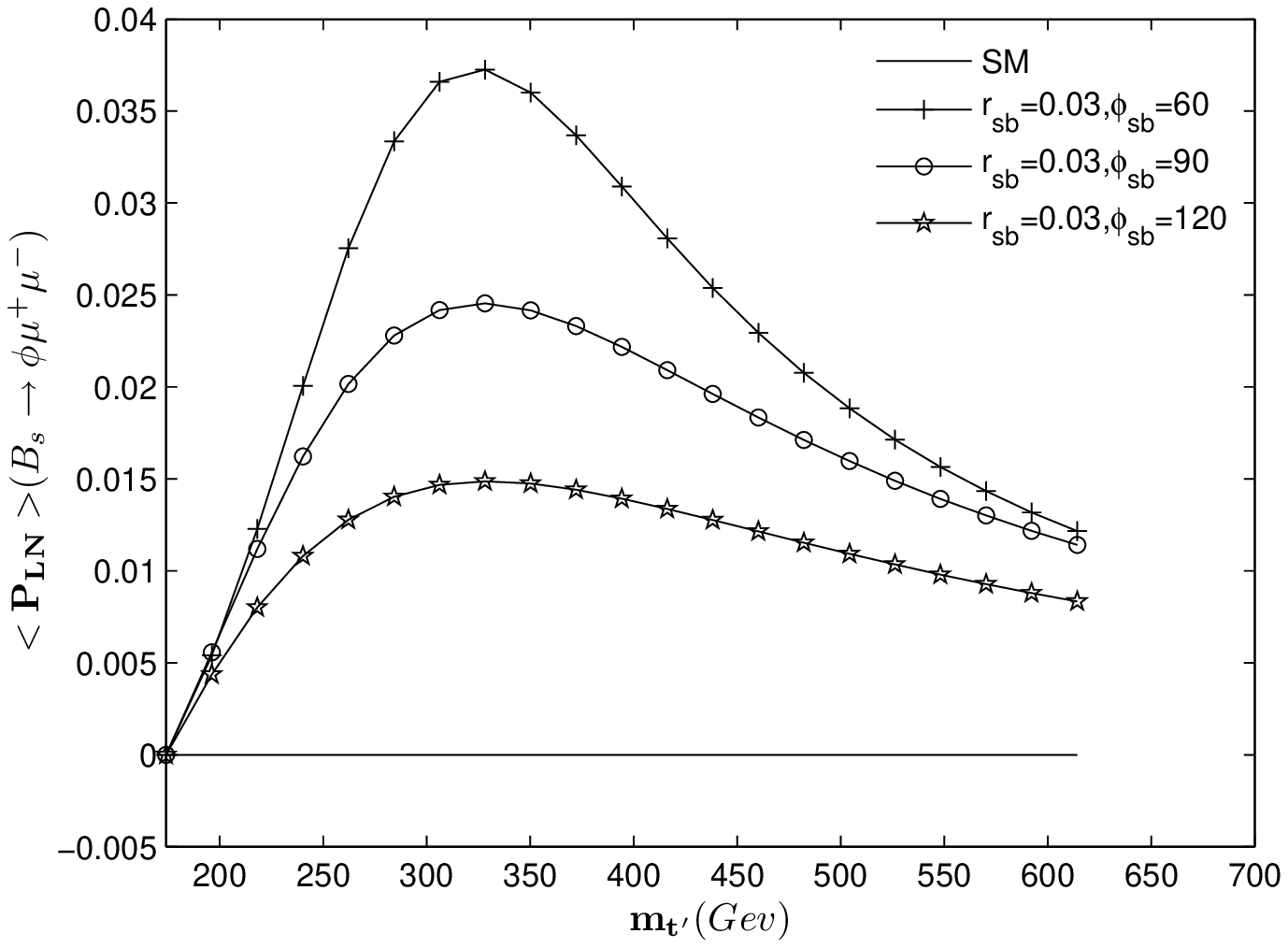}
        \includegraphics[height=2.1in]{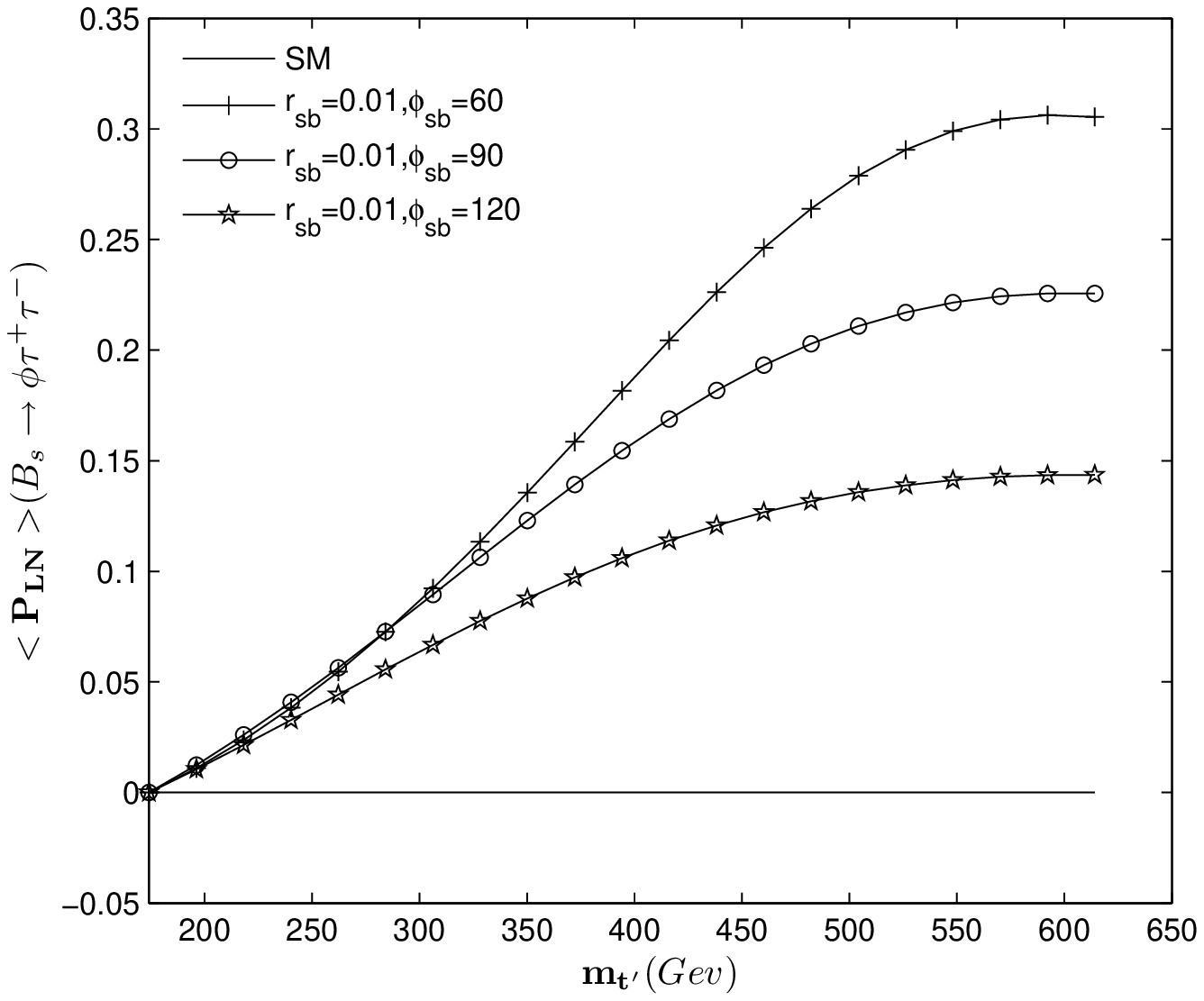}
        \includegraphics[height=2.1in]{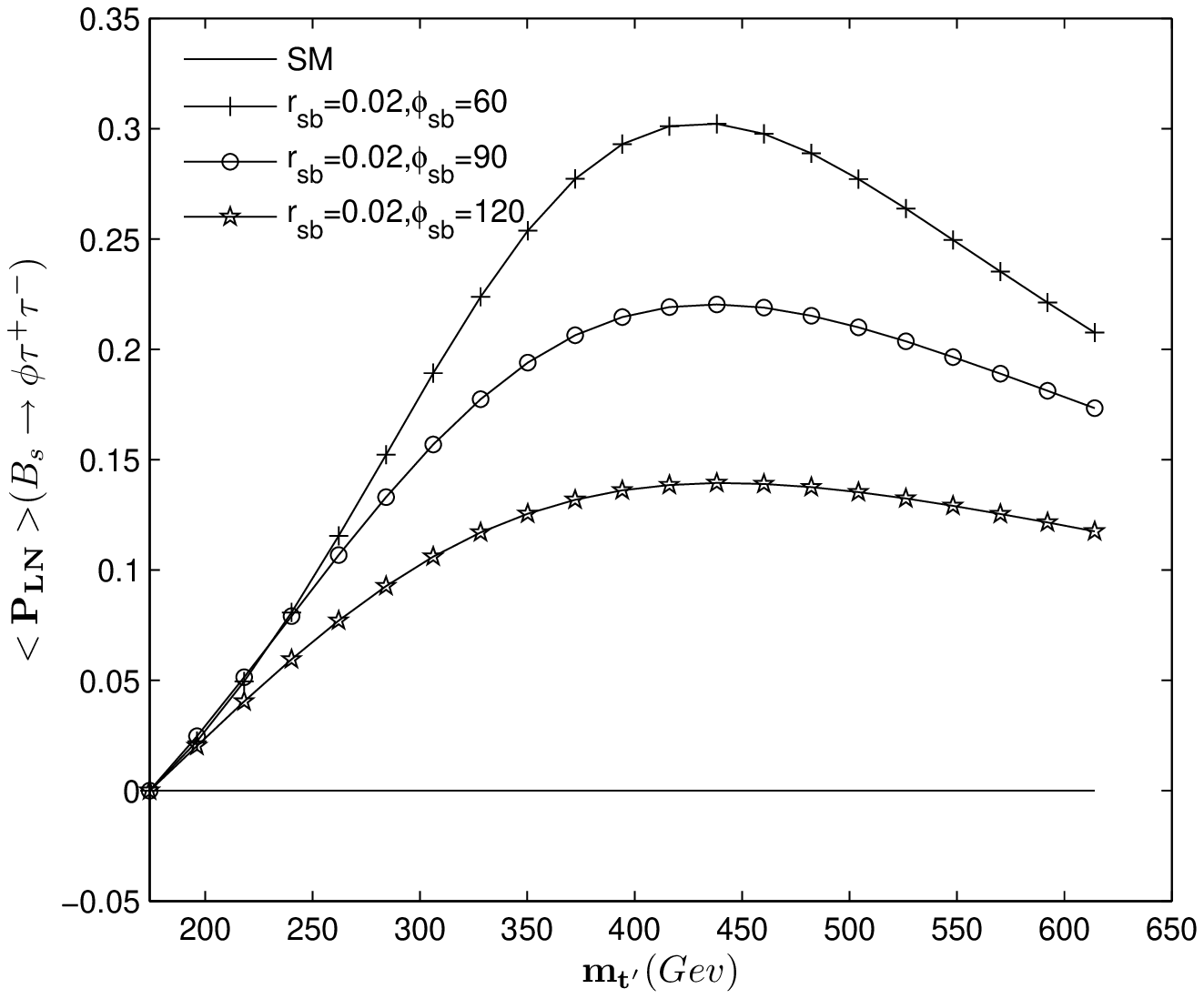}
         \includegraphics[height=2.1in]{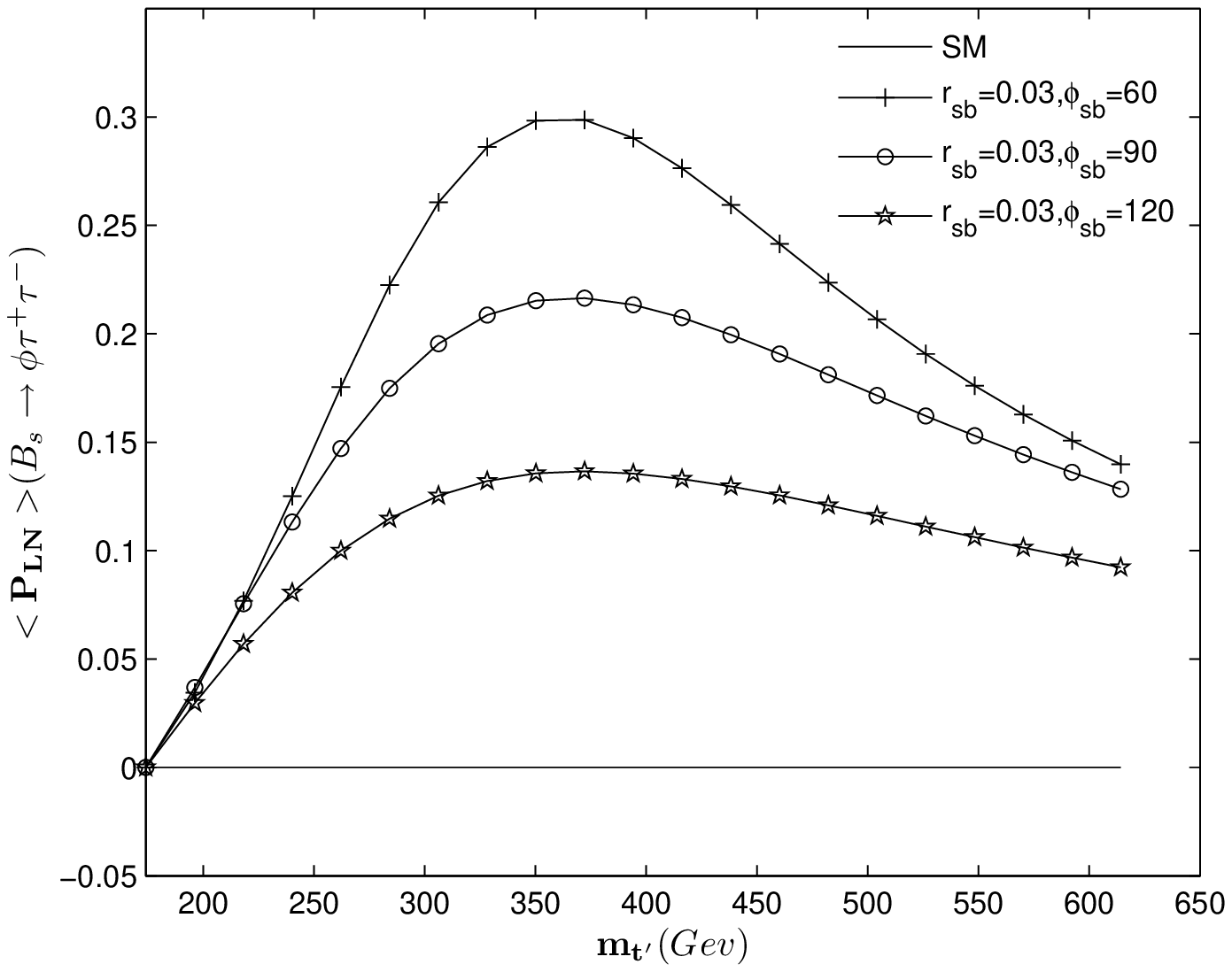}
                       \caption{The dependence of the $\lla P_{LN}\rra$  on the fourth-generation quark mass
$m_{t'}$ for three different values of
 $\phi_{sb}=\{60^\circ,~ 90^\circ, ~120^\circ\}$ and $r_{sb}=\{0.01,~0.02,~0.03\}$ for the $\mu$ and $\tau$ channels.}
\label{PLN}
           \end{minipage} }

      \end{figure}

    \begin{figure}
  \centering
  \setlength{\fboxrule}{2pt}
 \fbox{ \begin{minipage}{6 in}
        \centering
\includegraphics[height=2.1in]{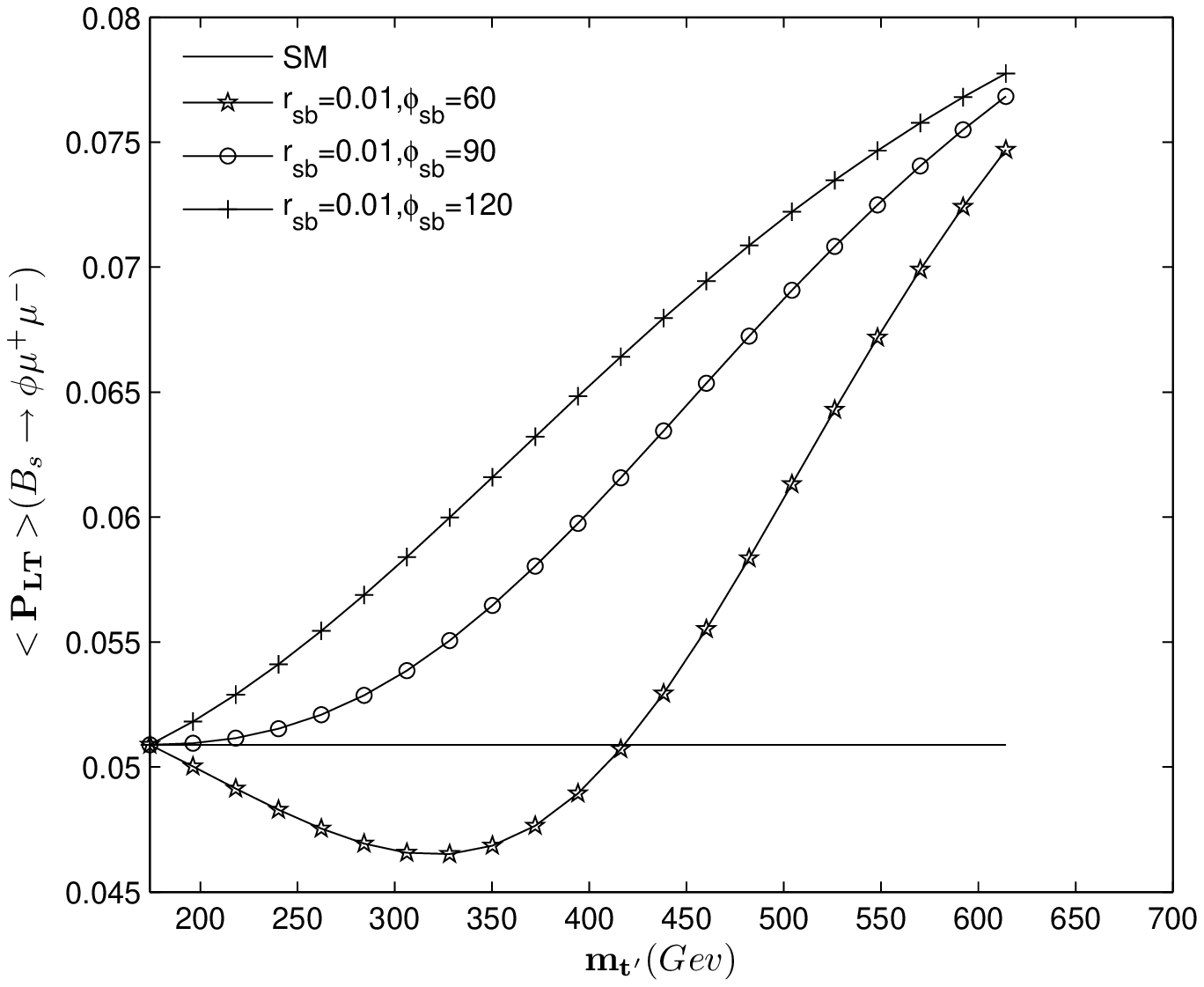}
        \includegraphics[height=2.1in]{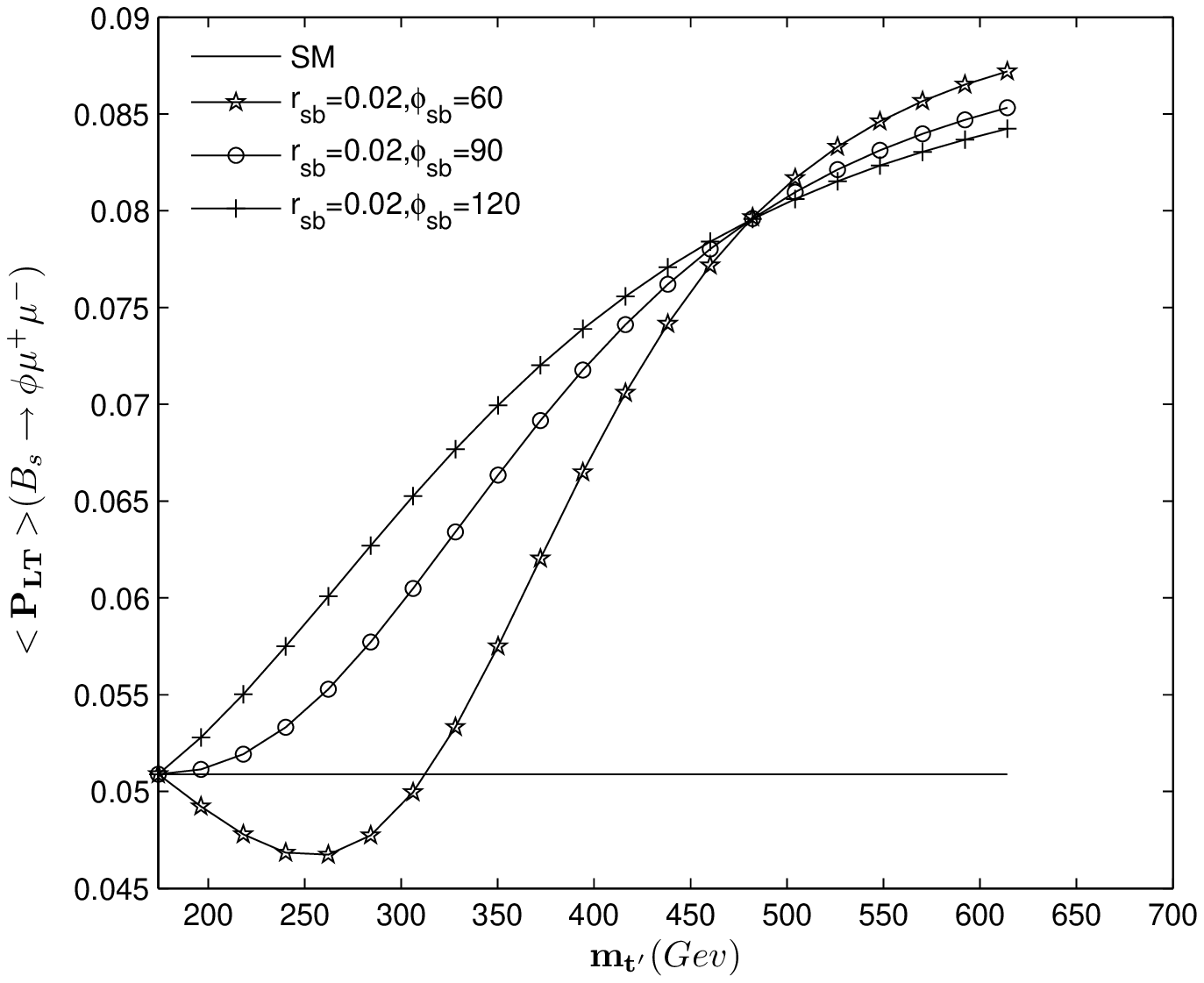}
         \includegraphics[height=2.1in]{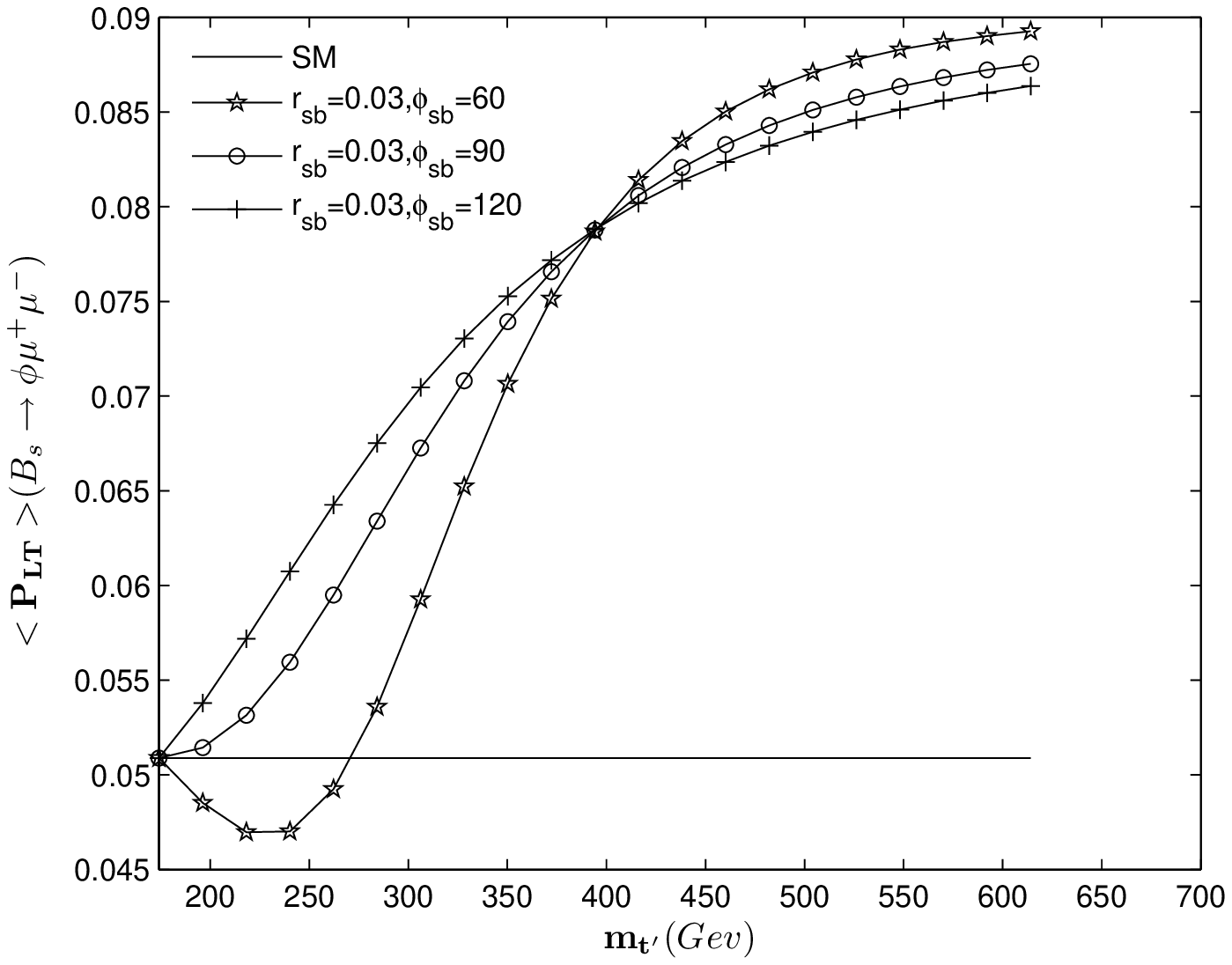}
        \includegraphics[height=2.1in]{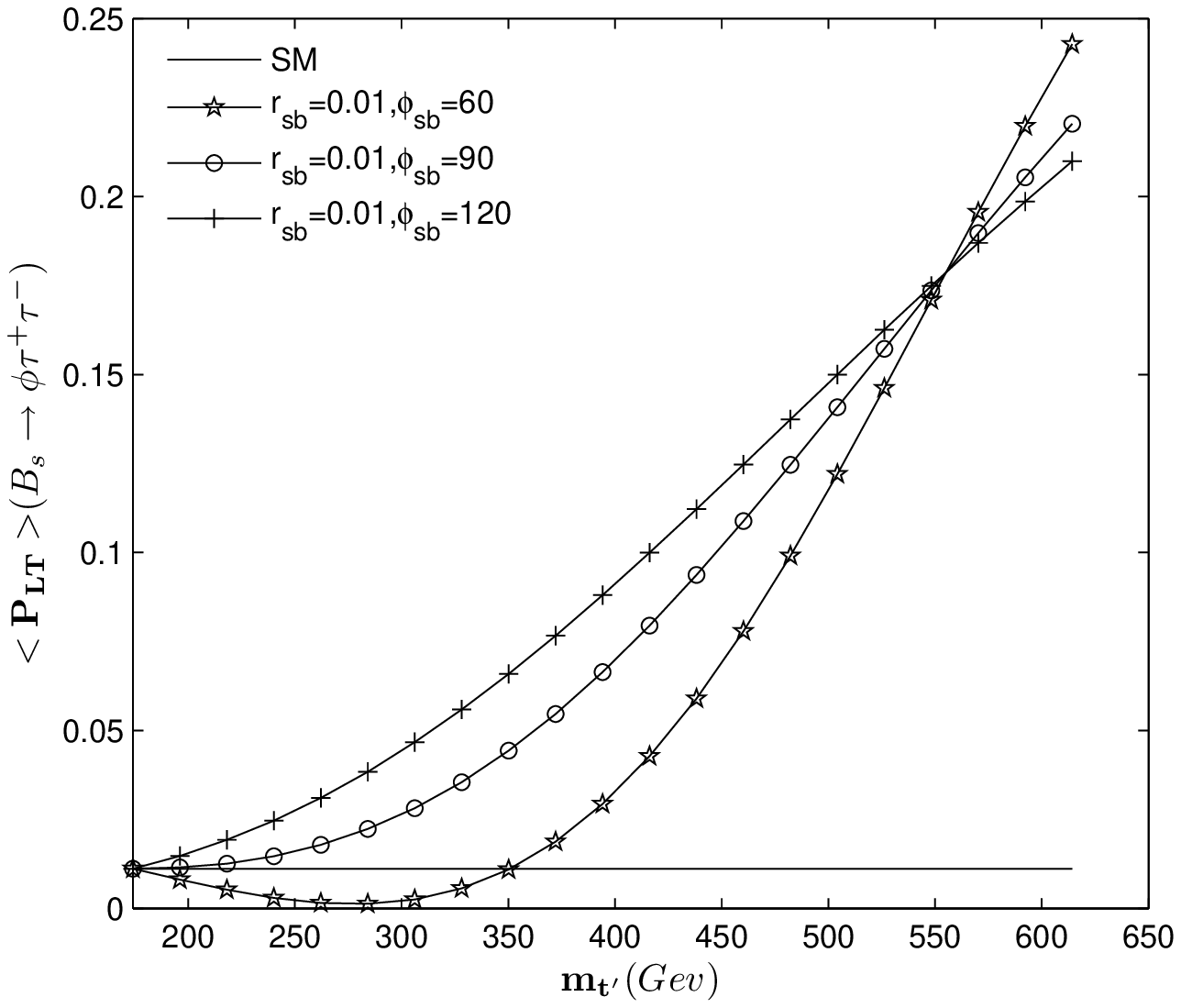}
        \includegraphics[height=2.1in]{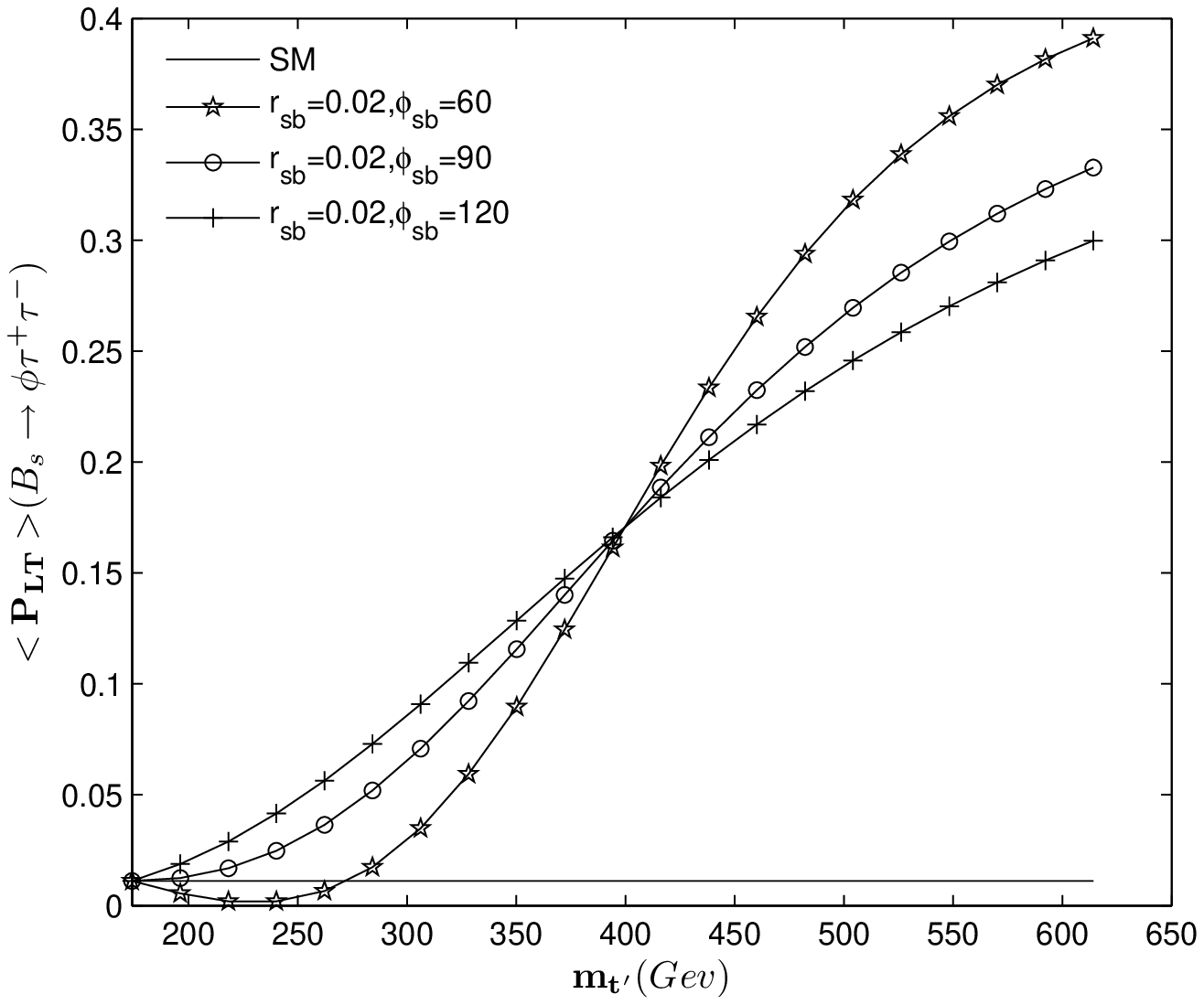}
         \includegraphics[height=2.1in]{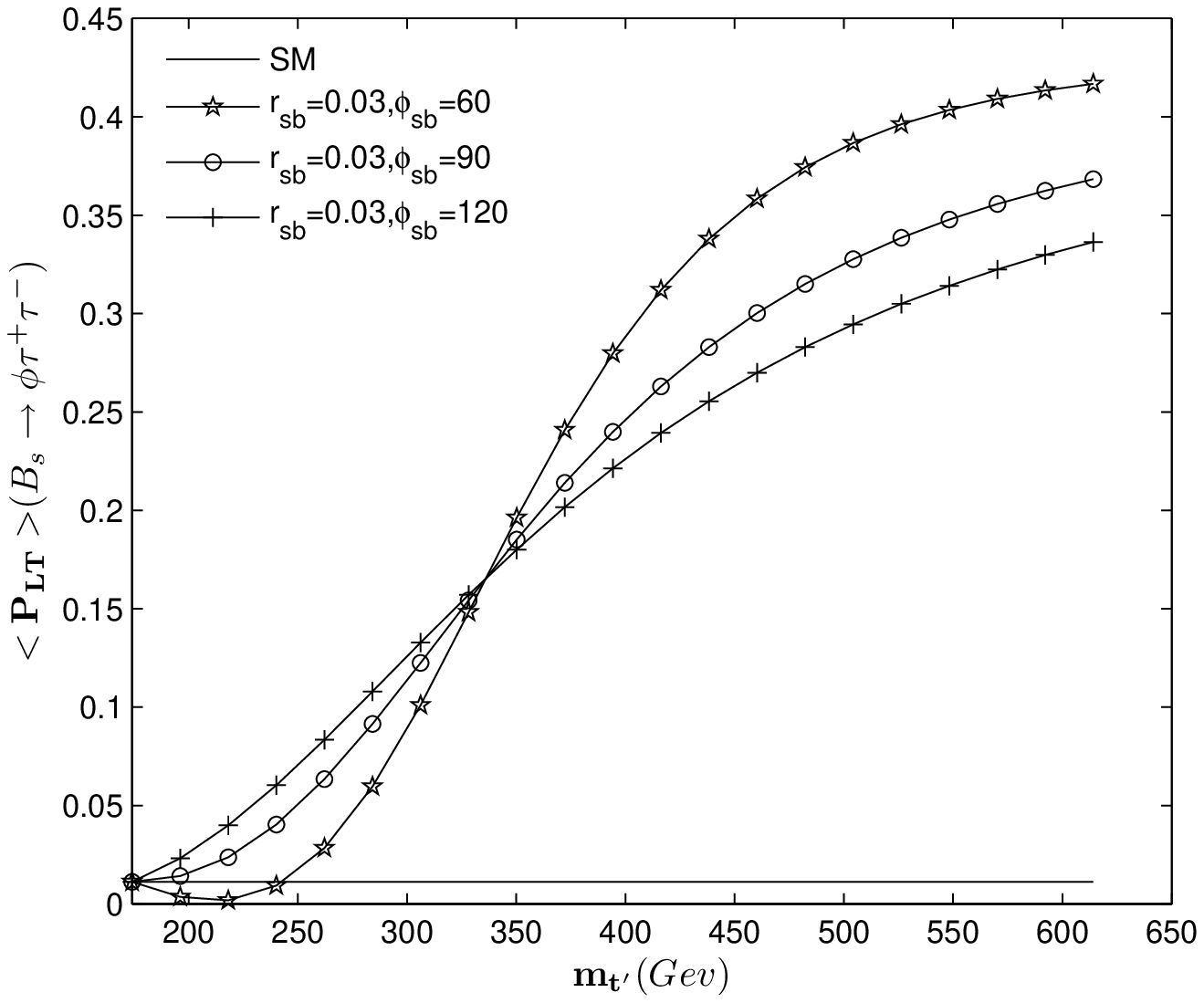}
                                 \caption{The dependence of the $\lla P_{LT}\rra$  on the fourth-generation quark mass
$m_{t'}$ for three different values of
 $\phi_{sb}=\{60^\circ,~ 90^\circ, ~120^\circ\}$ and $r_{sb}=\{0.01,~0.02,~0.03\}$ for the $\mu$ and $\tau$ channels.}
\label{PLT}
 \end{minipage} }
  \end{figure}
\begin{figure}
  \centering
  \setlength{\fboxrule}{2pt}
 \fbox{ \begin{minipage}{6 in}
        \centering
     \includegraphics[height=2.1in]{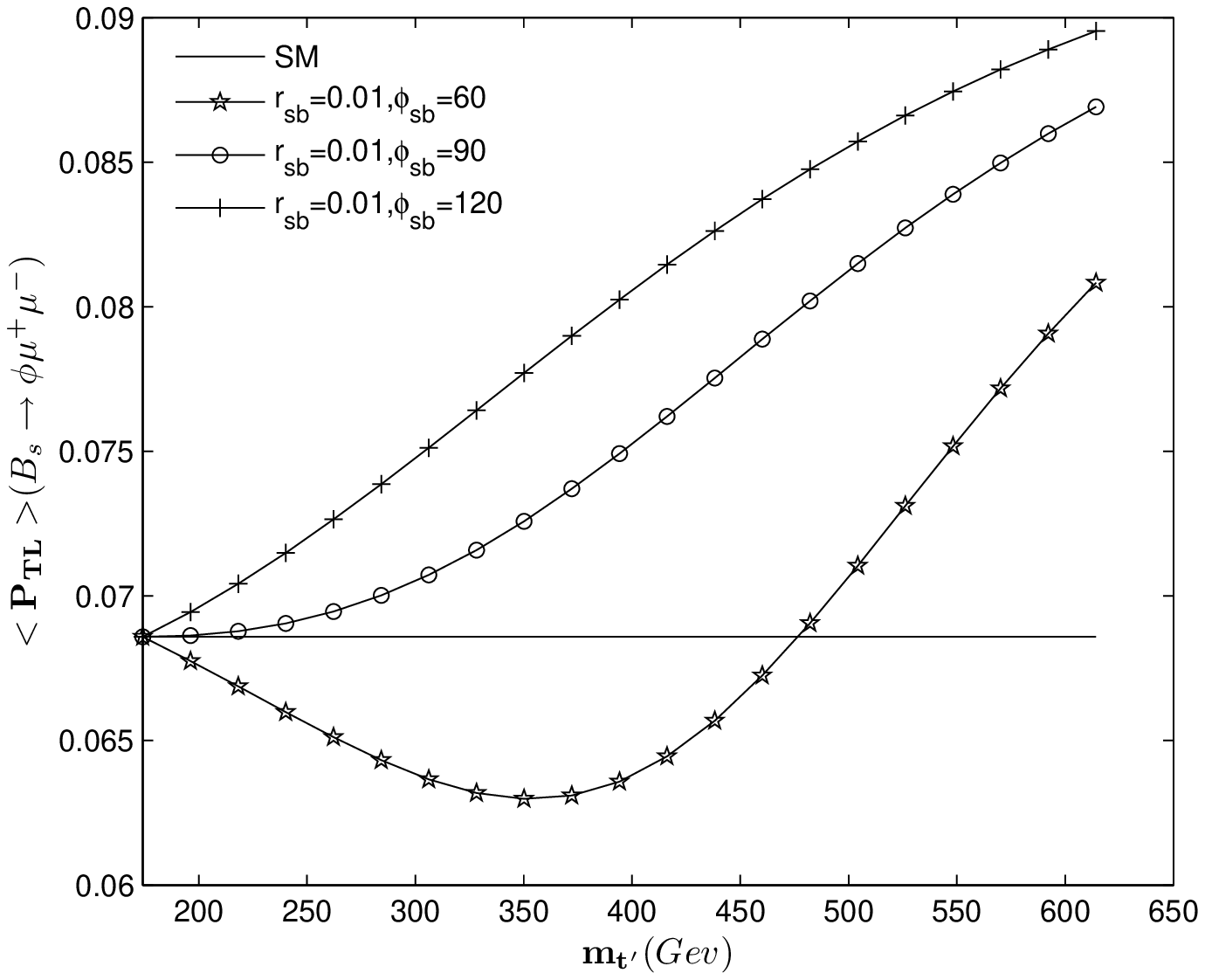}
        \includegraphics[height=2.1in]{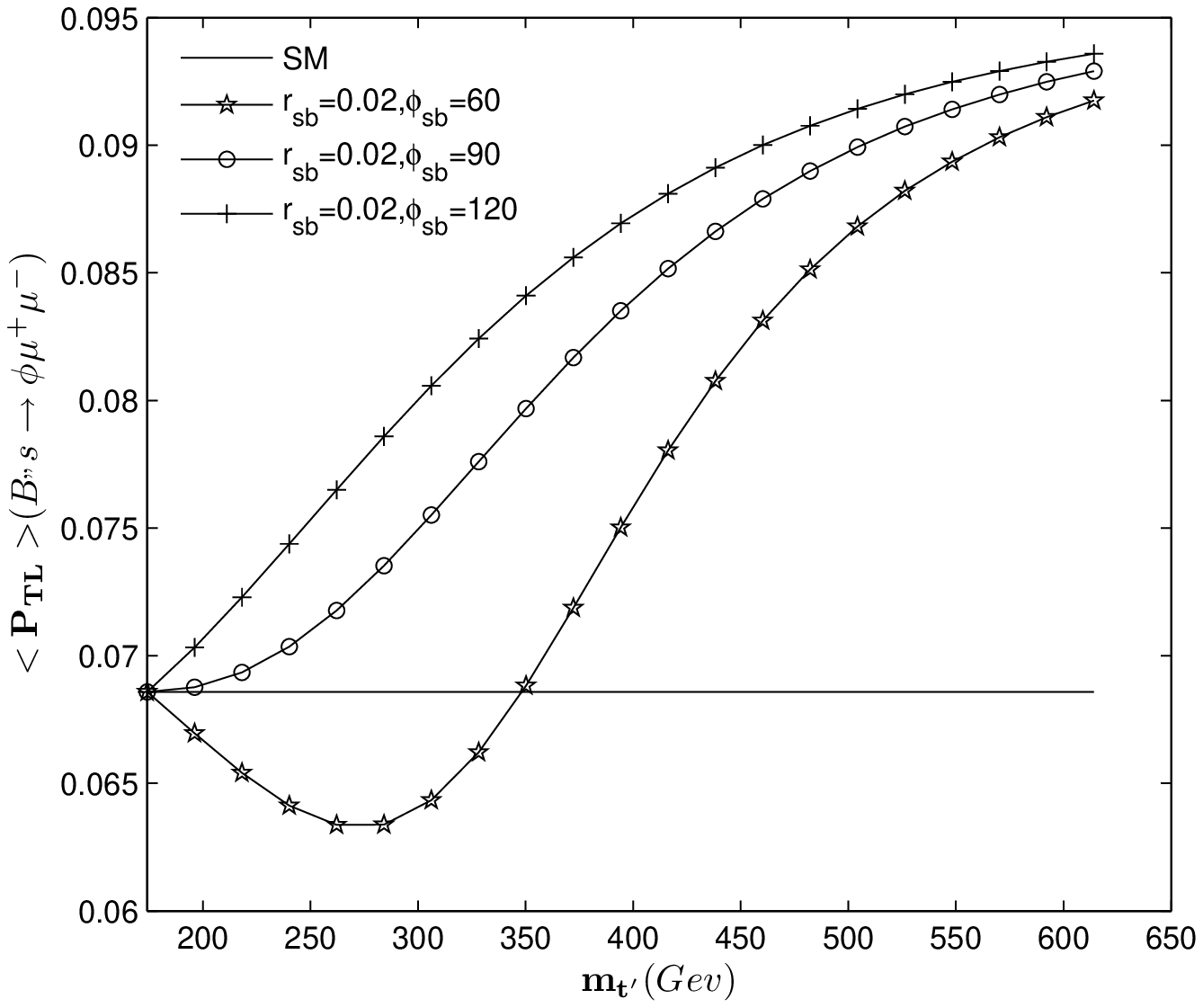}
         \includegraphics[height=2.1in]{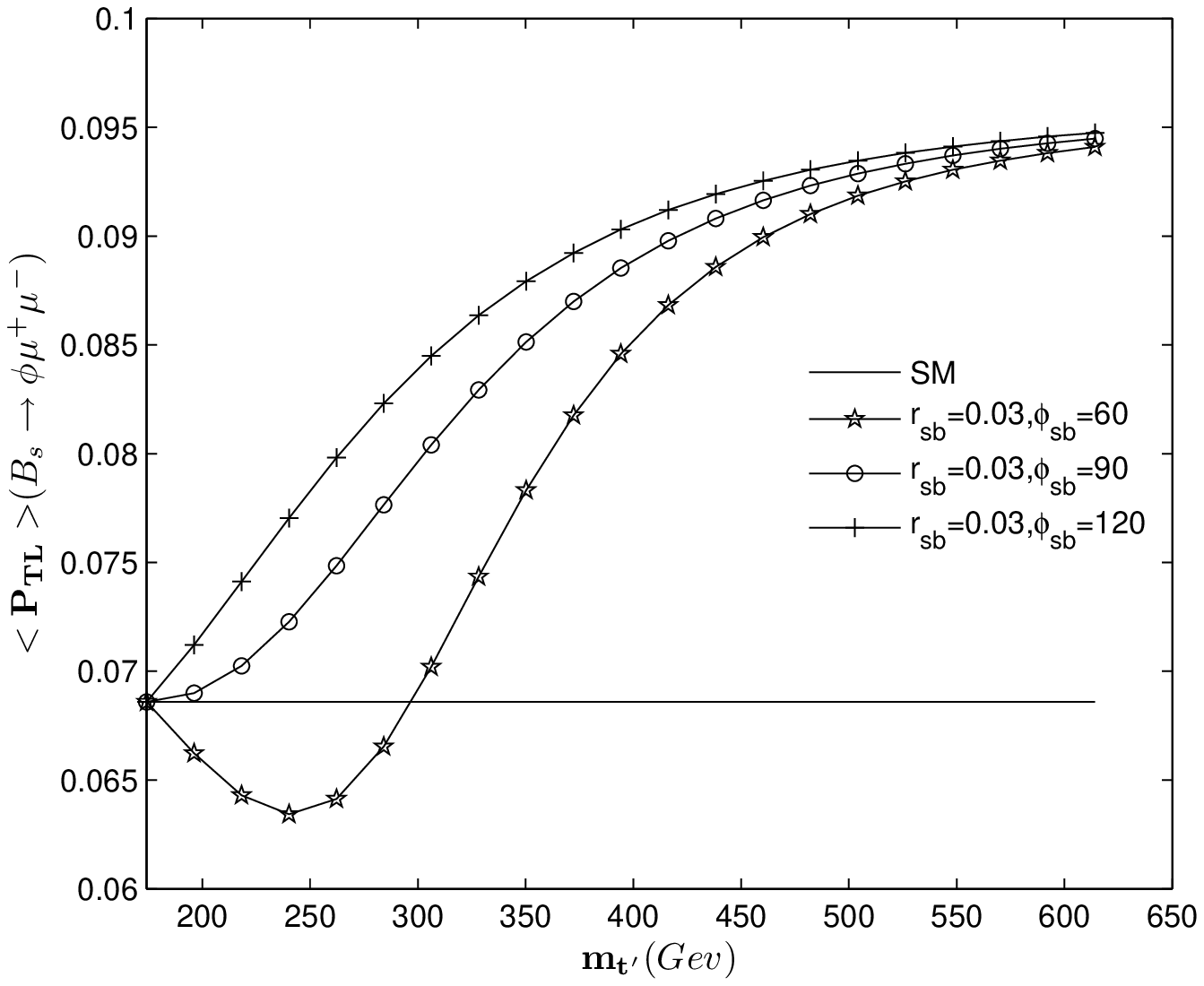}
        \includegraphics[height=2.1in]{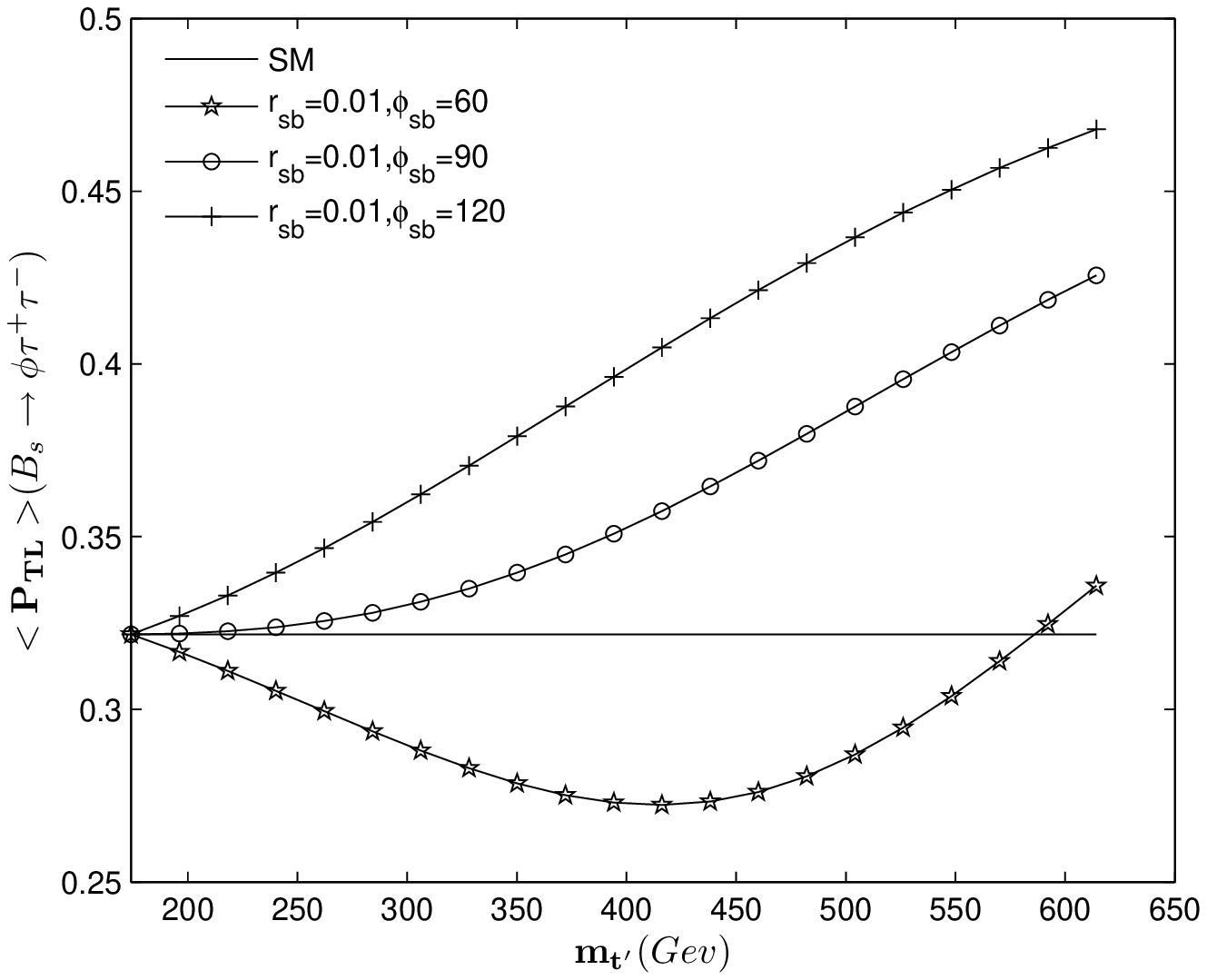}
        \includegraphics[height=2.1in]{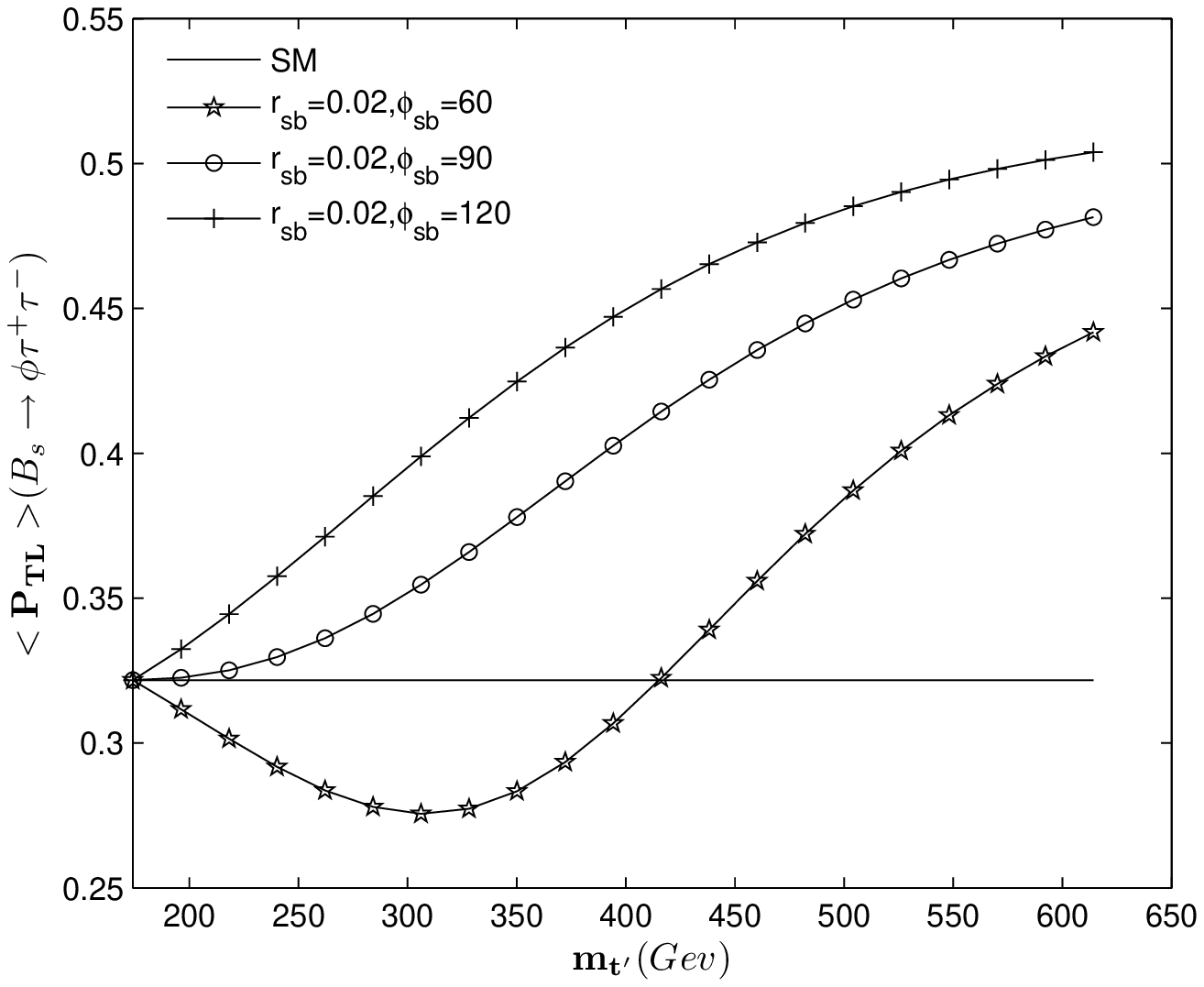}
         \includegraphics[height=2.1in]{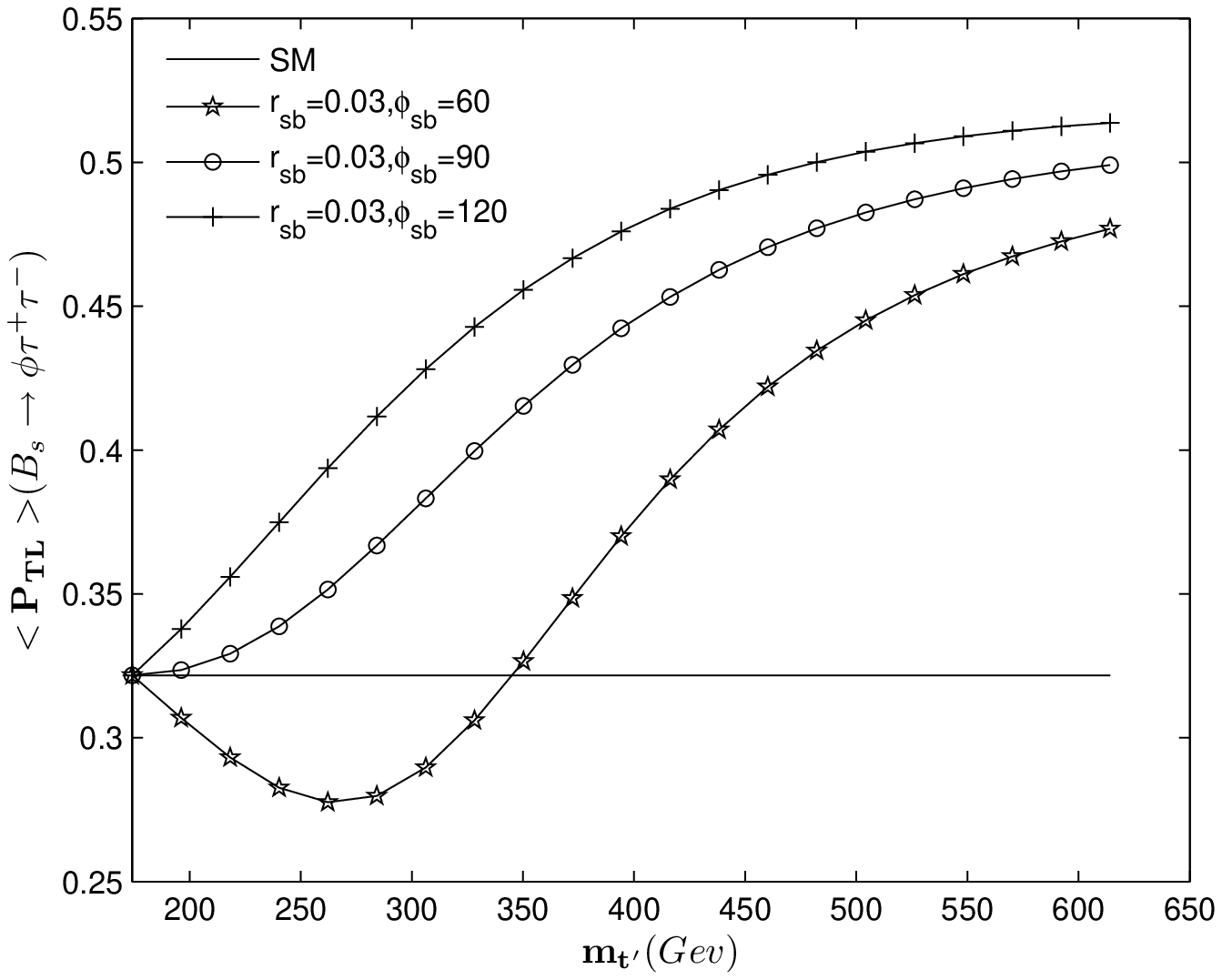}
                \caption{The dependence of the $\lla P_{TL}\rra$  on the fourth-generation quark mass
$m_{t'}$ for three different values of
 $\phi_{sb}=\{60^\circ,~ 90^\circ, ~120^\circ\}$ and $r_{sb}=\{0.01,~0.02,~0.03\}$ for the $\mu$ and $\tau$ channels. }
\label{PTL}
                  \end{minipage} }

  \end{figure}
\begin{figure}
  \centering
  \setlength{\fboxrule}{2pt}
 \fbox{ \begin{minipage}{6 in}
        \centering
     \includegraphics[height=2.1in]{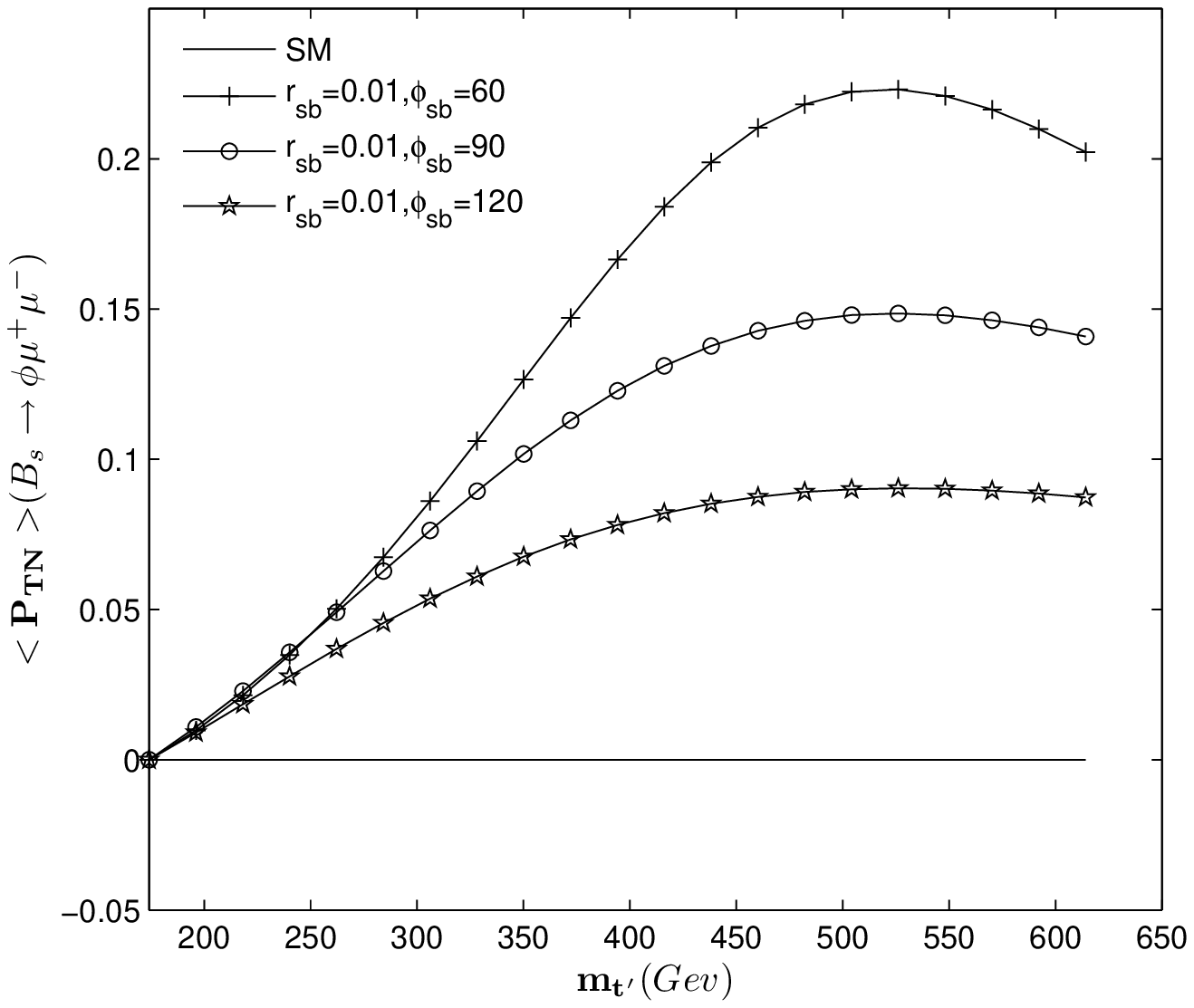}
        \includegraphics[height=2.1in]{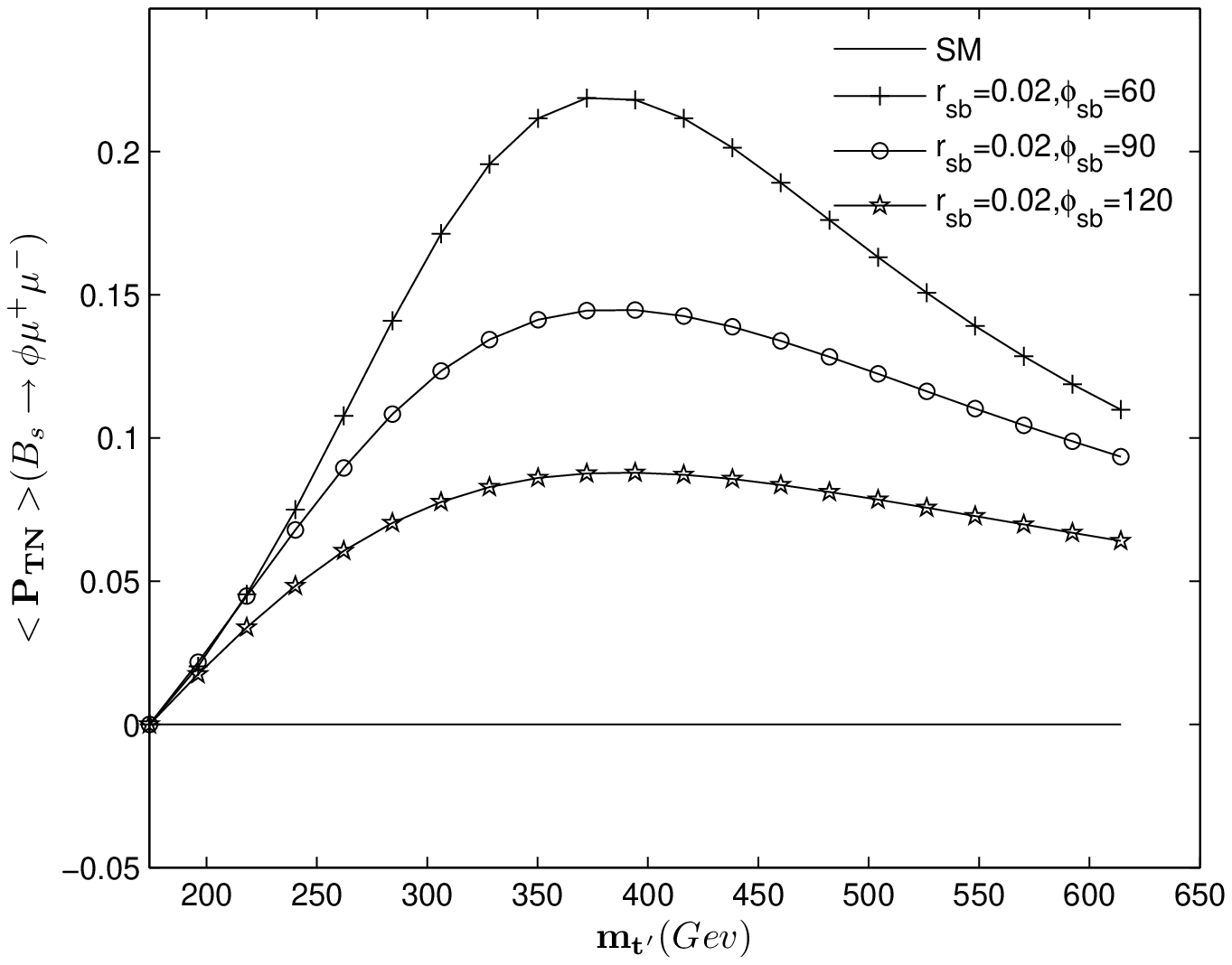}
         \includegraphics[height=2.1in]{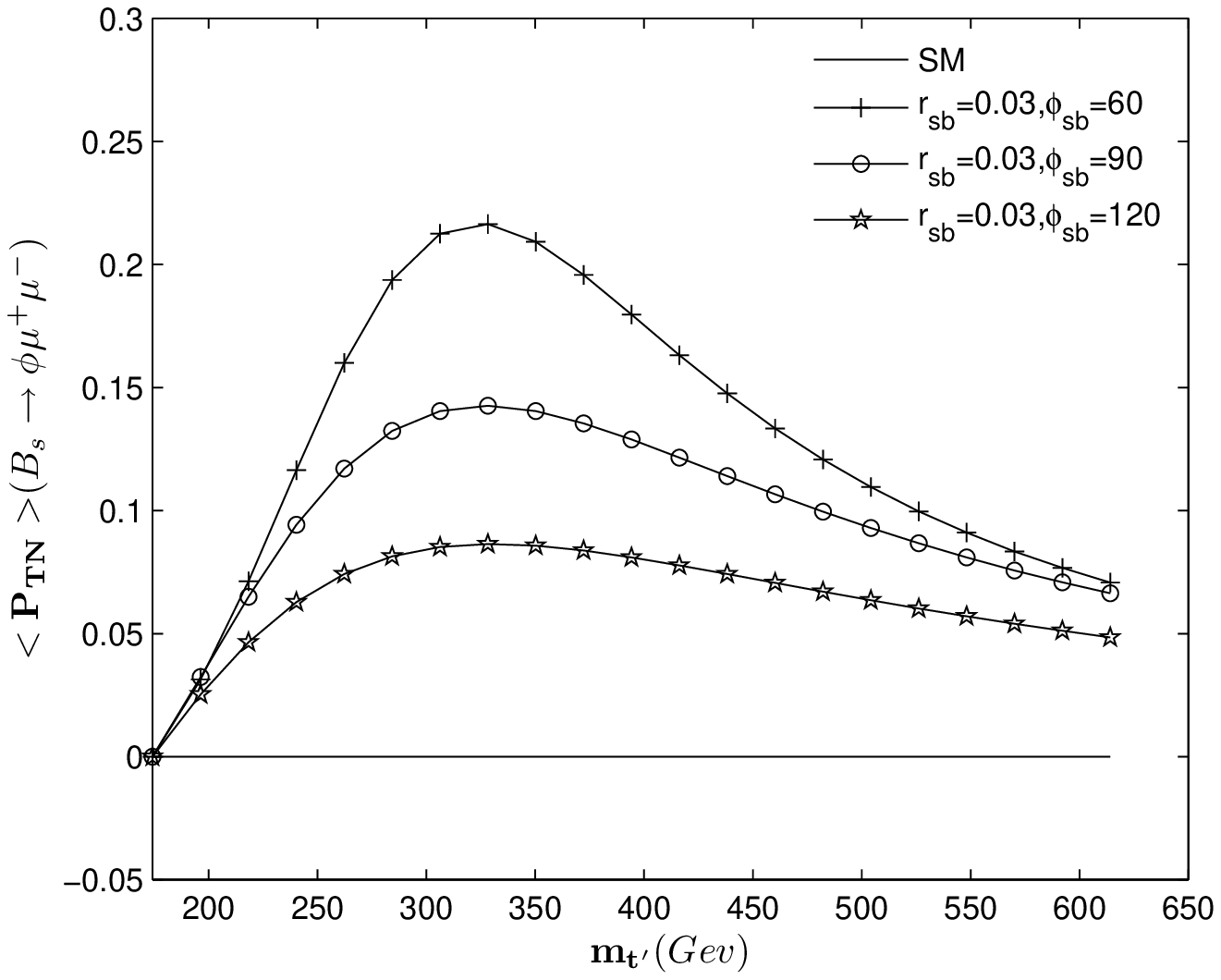}
        \includegraphics[height=2.1in]{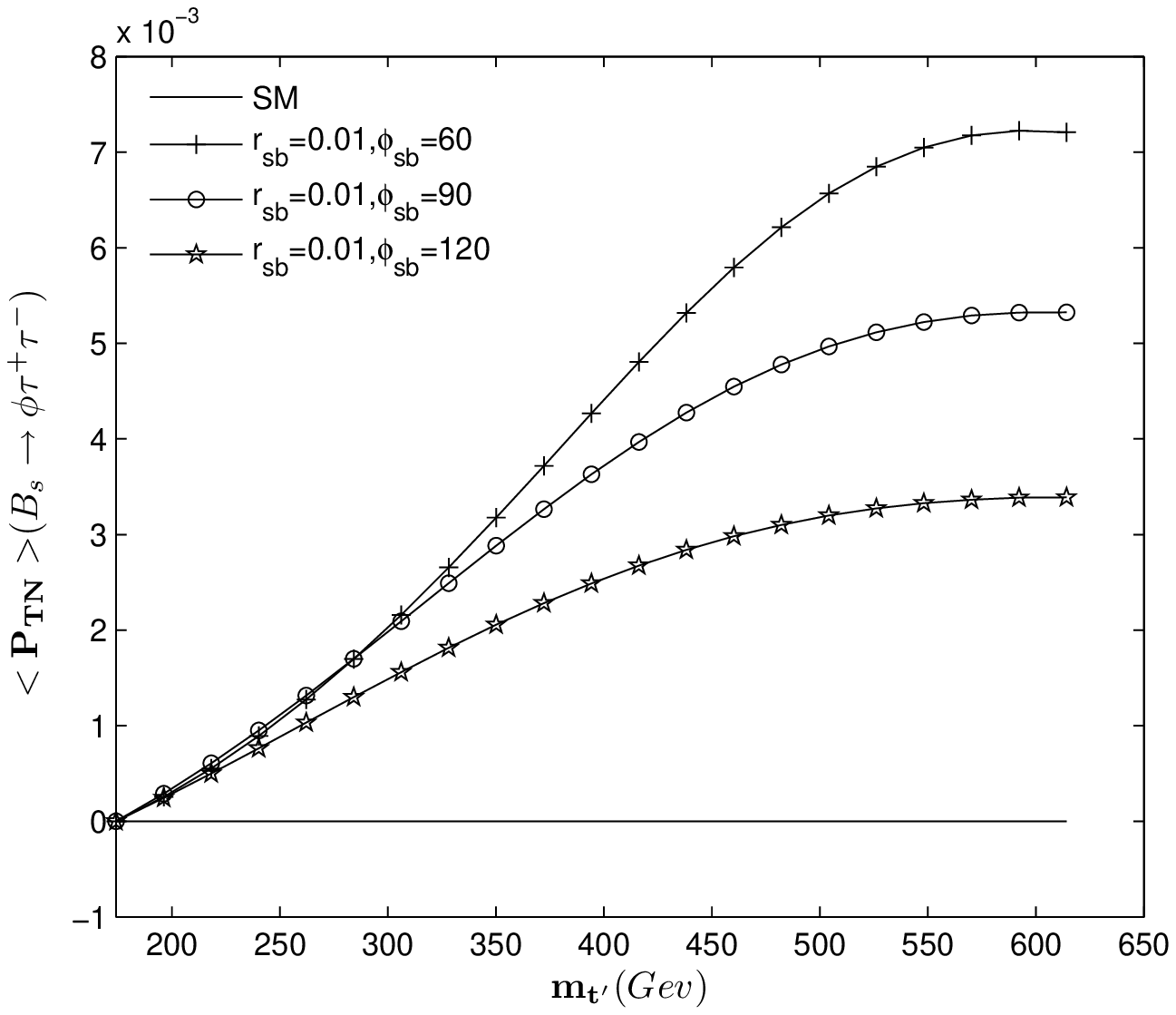}
        \includegraphics[height=2.1in]{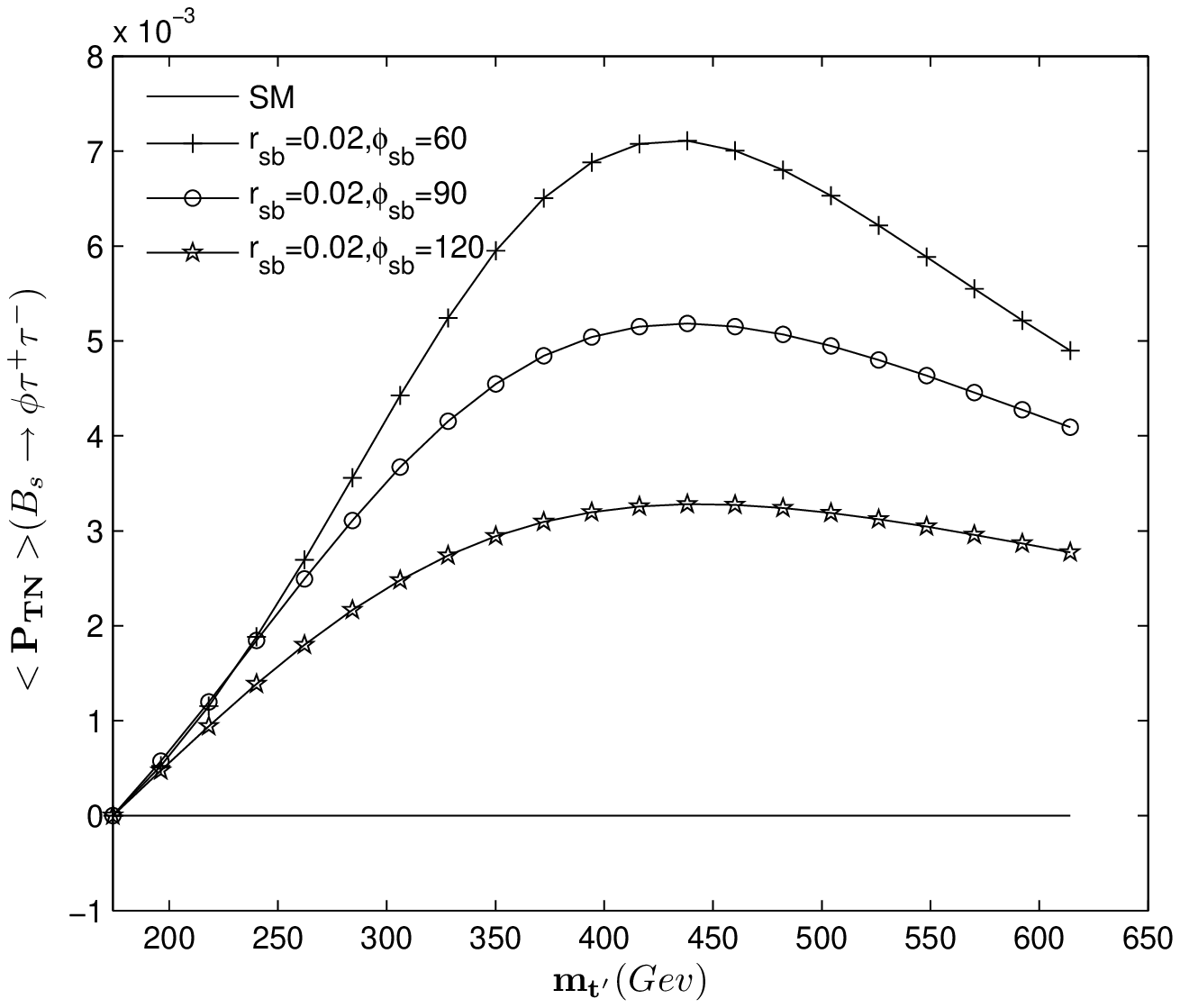}
         \includegraphics[height=2.1in]{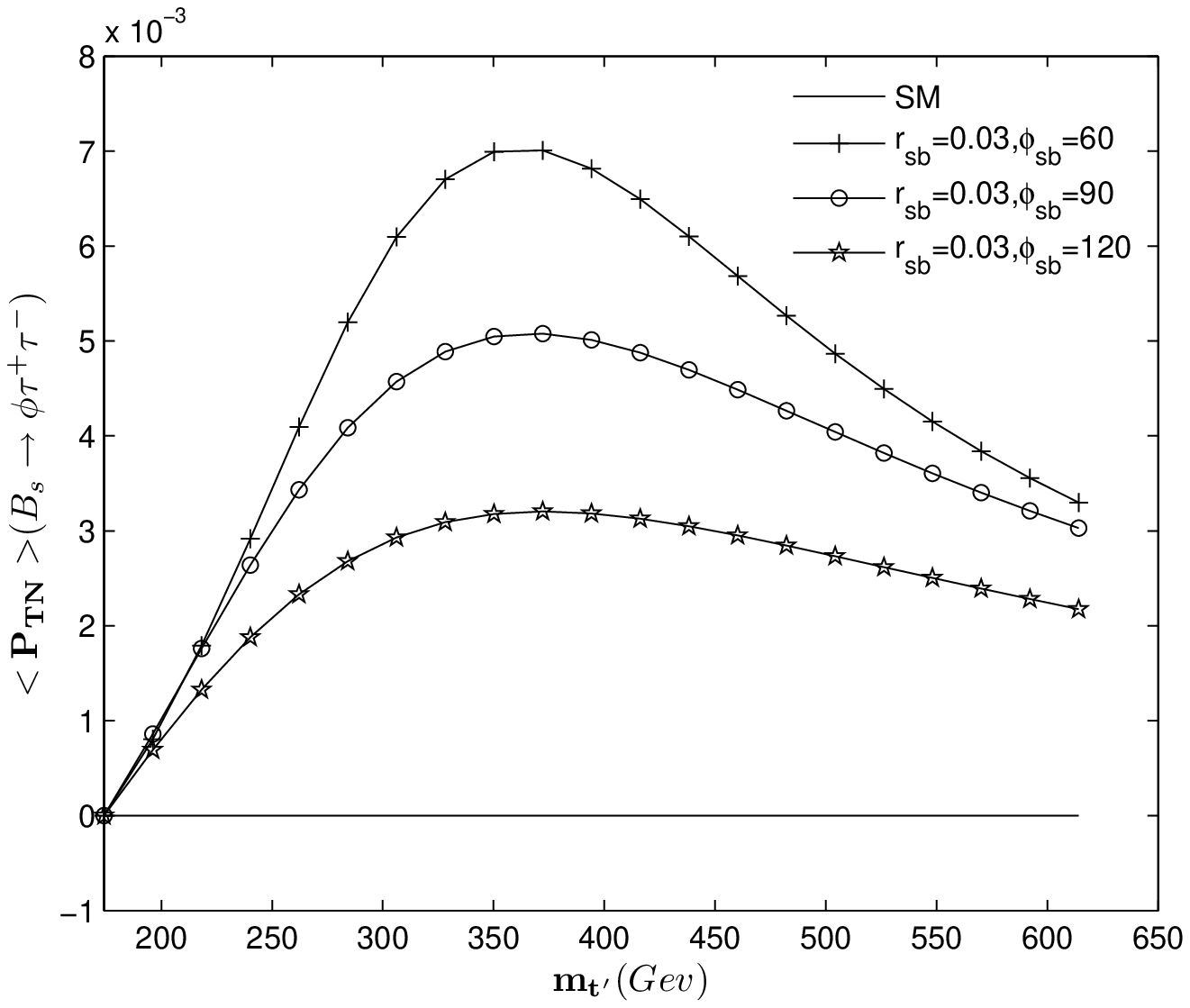}
                \caption{The dependence of the $\lla P_{TN}\rra$  on the fourth-generation quark mass
$m_{t'}$ for three different values of
 $\phi_{sb}=\{60^\circ,~ 90^\circ, ~120^\circ\}$ and $r_{sb}=\{0.01,~0.02,~0.03\}$ for the $\mu$ and $\tau$ channels. }
\label{PTN}
                  \end{minipage} }

  \end{figure}
  \begin{figure}
  \centering
  \setlength{\fboxrule}{2pt}
 \fbox{ \begin{minipage}{6 in}
        \centering
     \includegraphics[height=2.1in]{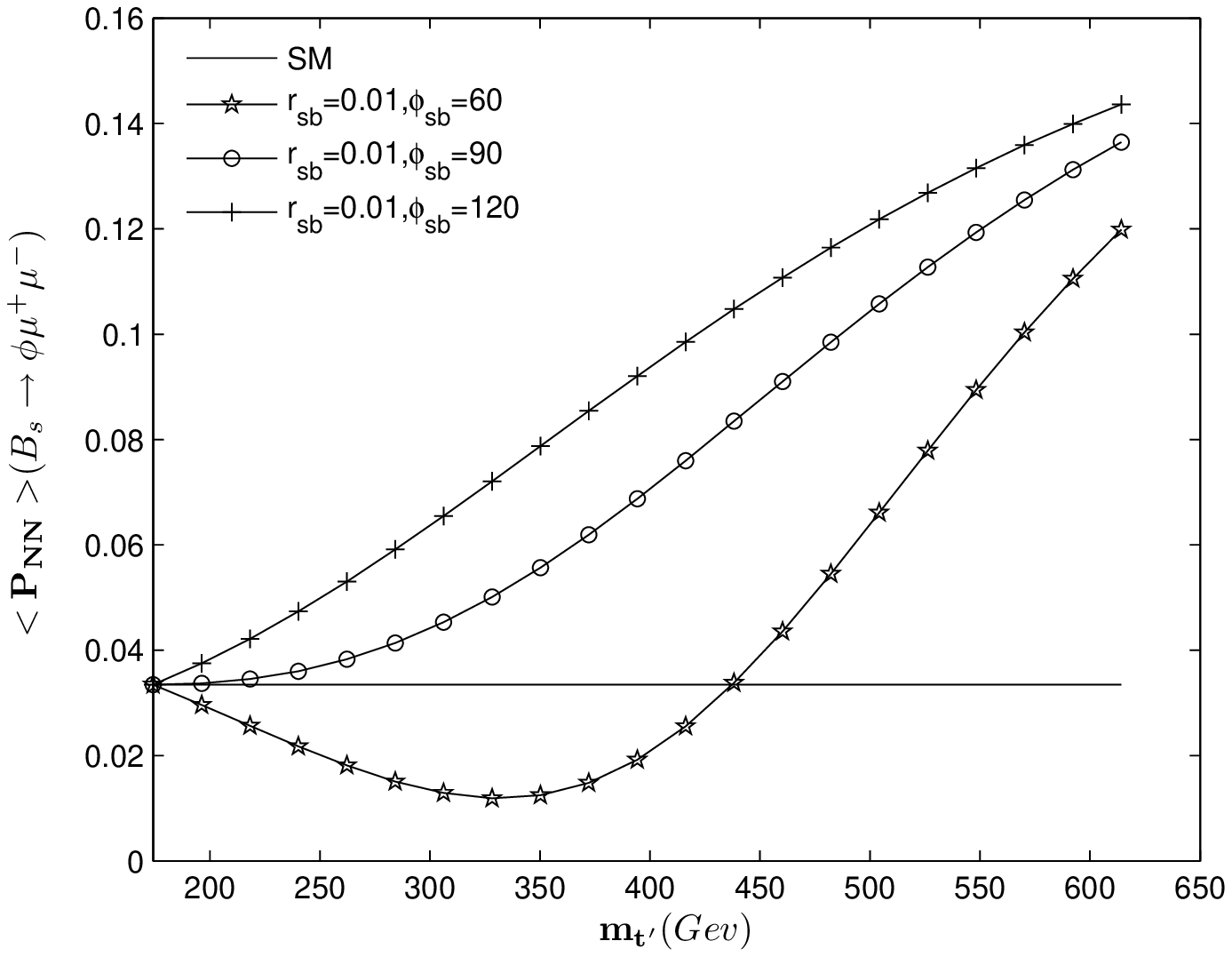}
        \includegraphics[height=2.1in]{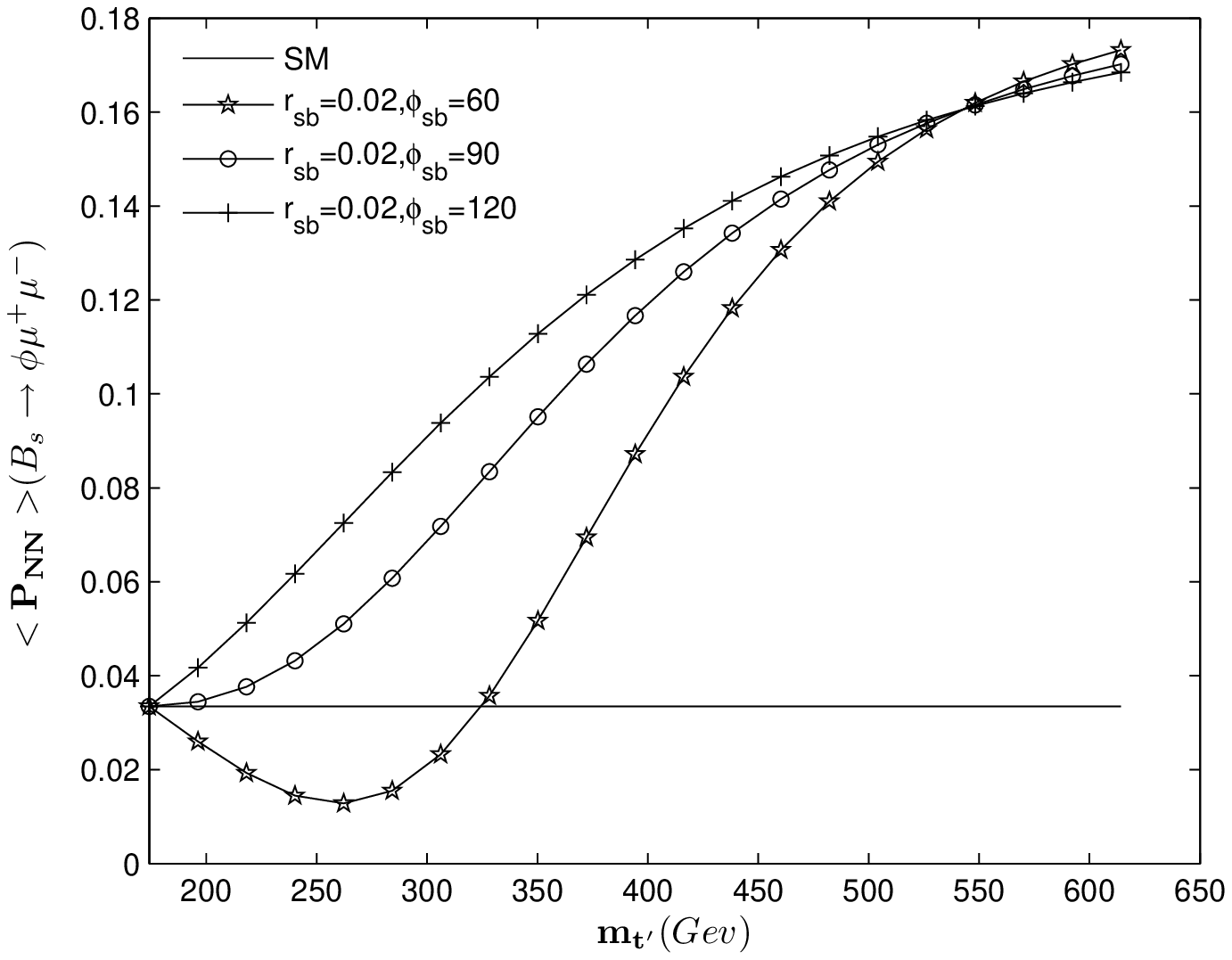}
         \includegraphics[height=2.1in]{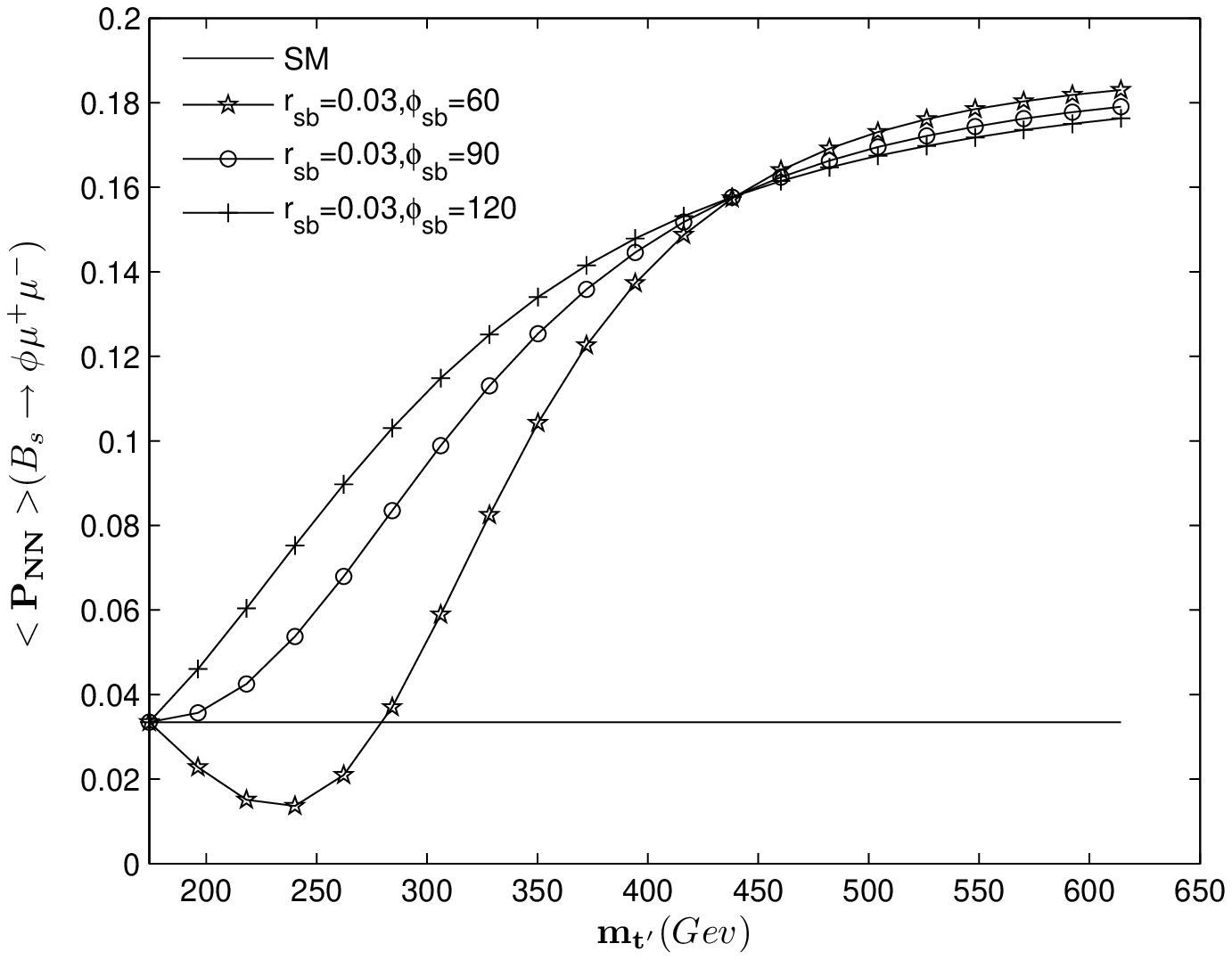}
        \includegraphics[height=2.1in]{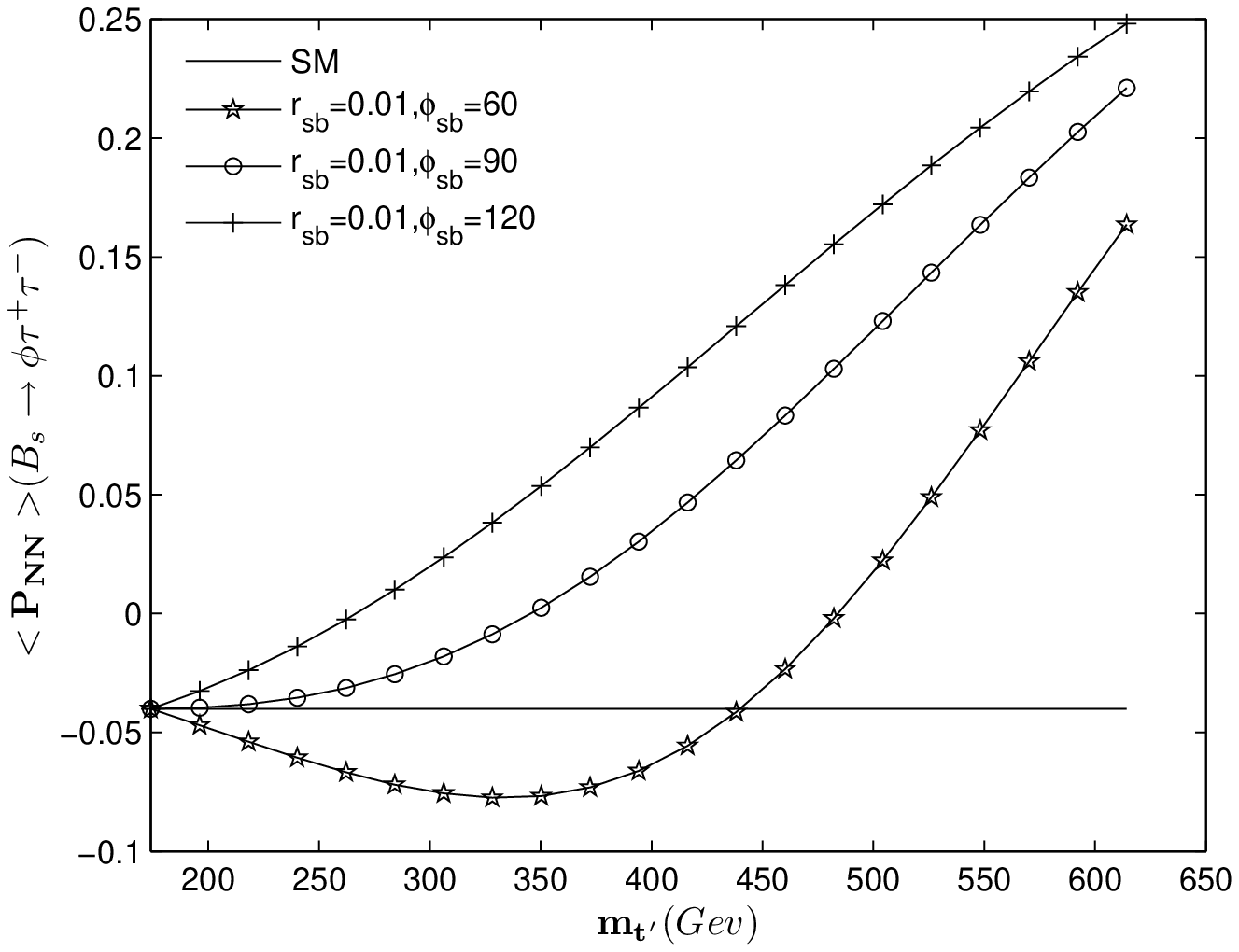}
        \includegraphics[height=2.1in]{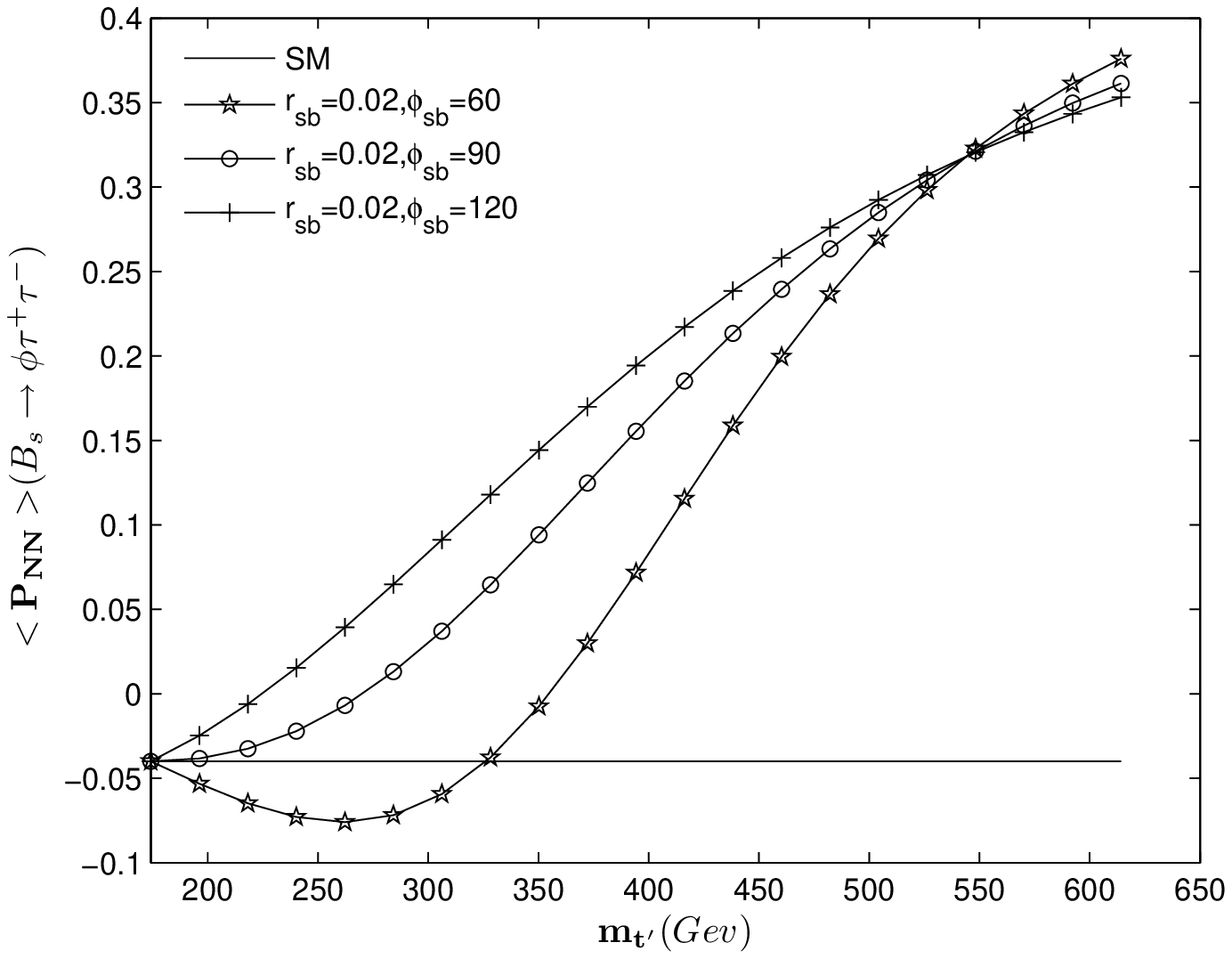}
         \includegraphics[height=2.1in]{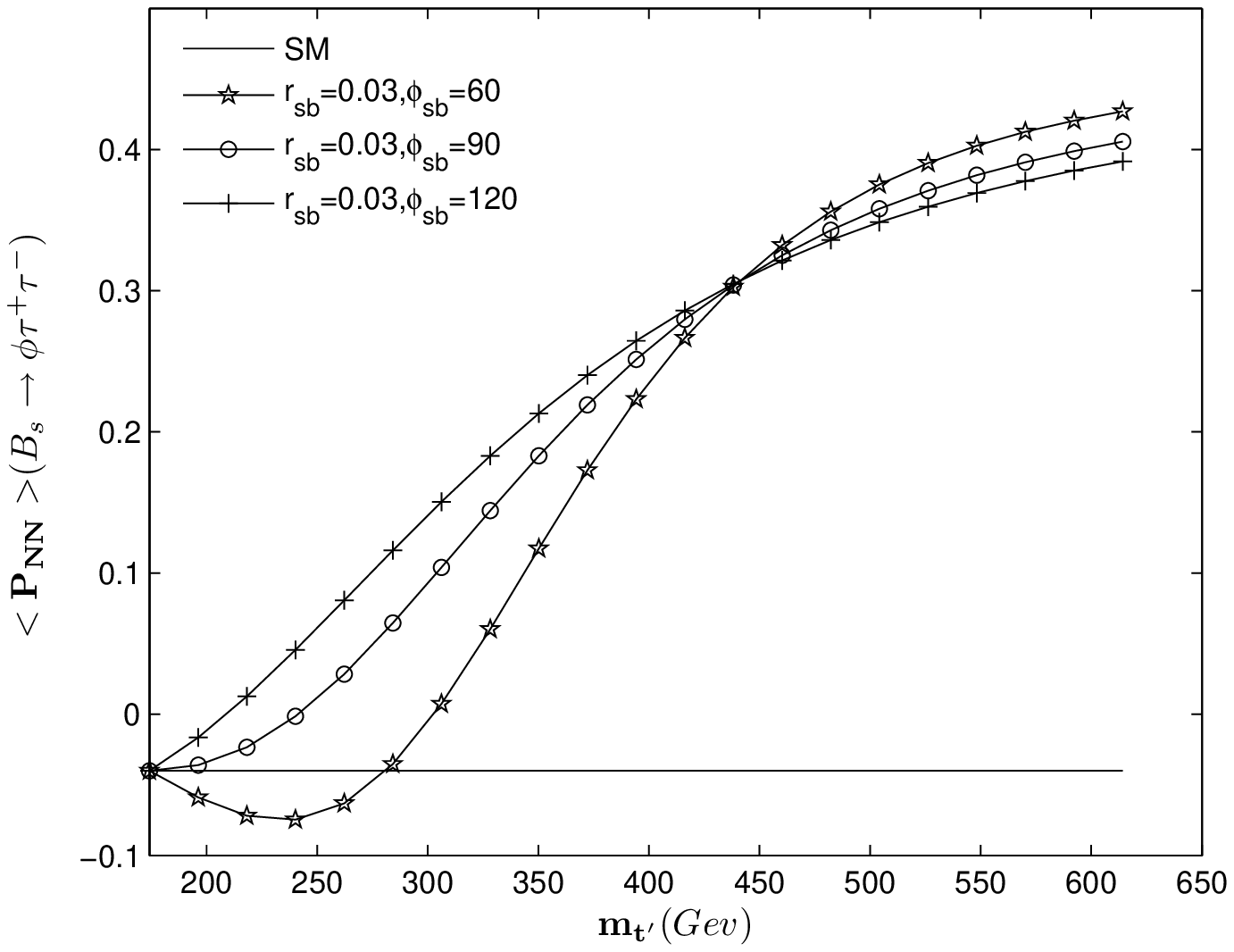}
                \caption{The dependence of the $\lla P_{NN}\rra$  on the fourth-generation quark mass
$m_{t'}$ for three different values of
 $\phi_{sb}=\{60^\circ,~ 90^\circ, ~120^\circ\}$ and $r_{sb}=\{0.01,~0.02,~0.03\}$ for the $\mu$ and $\tau$ channels. }
\label{PNN}
                  \end{minipage} }
 \end{figure}
\begin{figure}
  \centering
  \setlength{\fboxrule}{2pt}
 \fbox{ \begin{minipage}{6 in}
        \centering
        \includegraphics[height=2.1in]{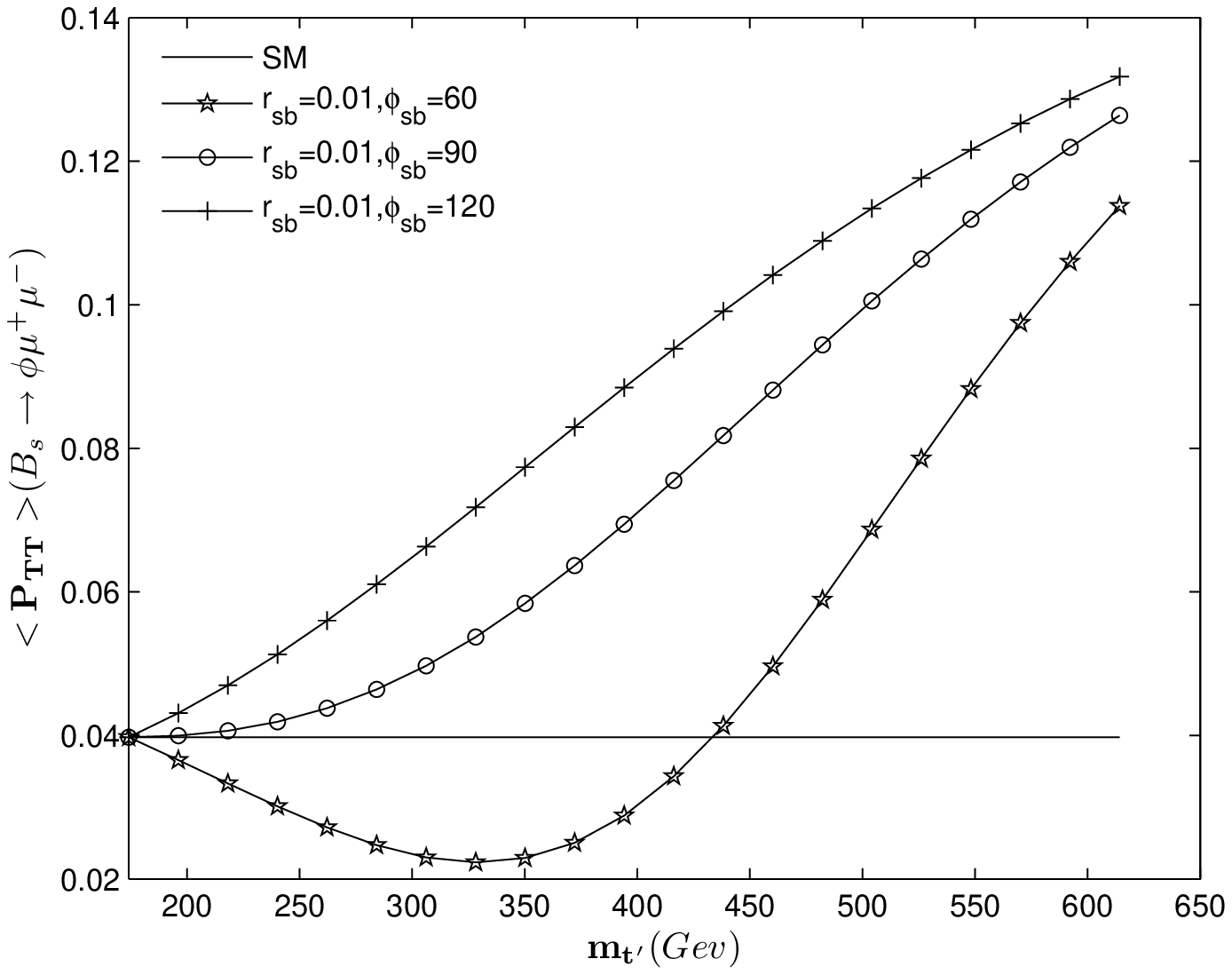}
        \includegraphics[height=2.1in]{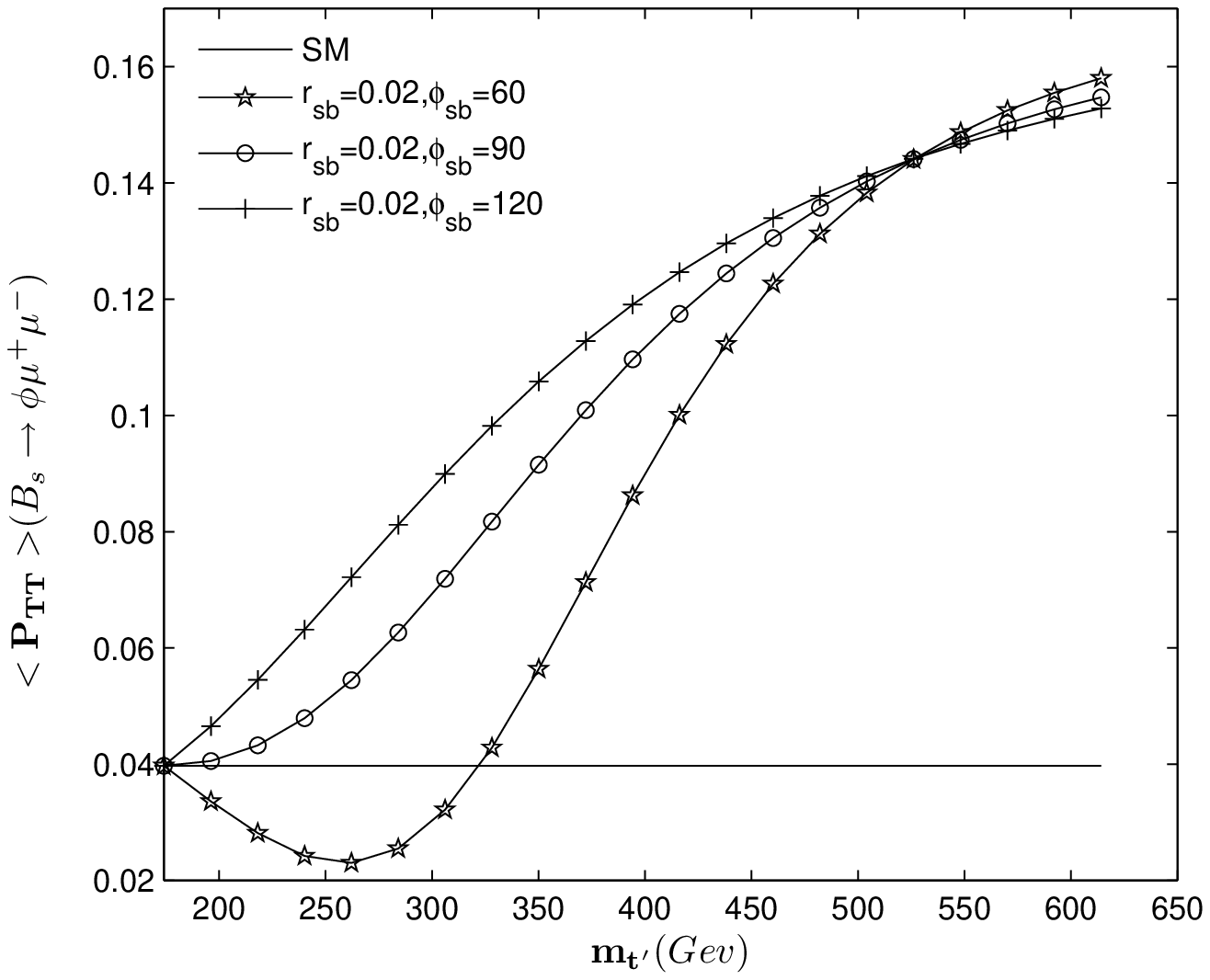}
        \includegraphics[height=2.1in]{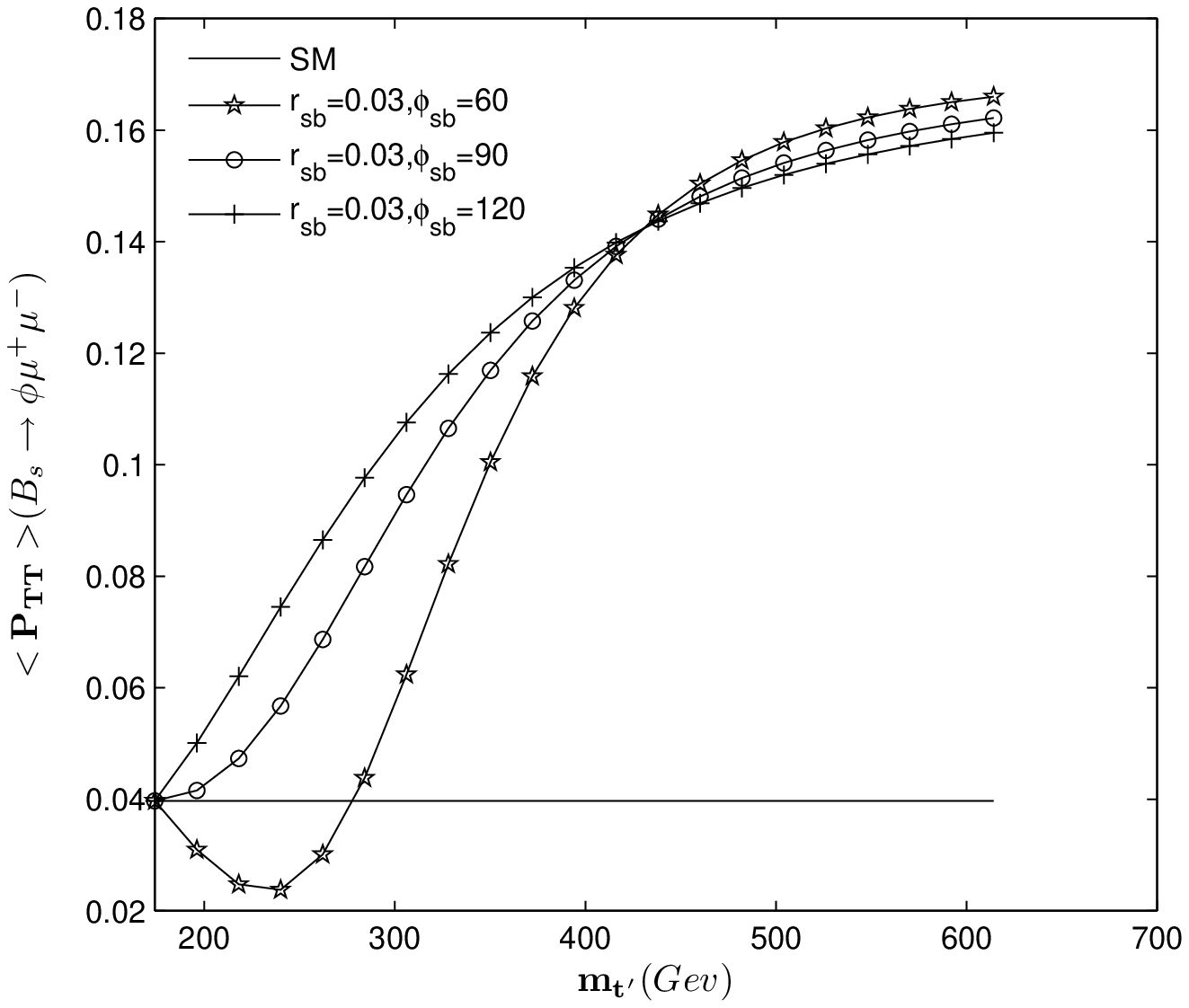}
        \includegraphics[height=2.1in]{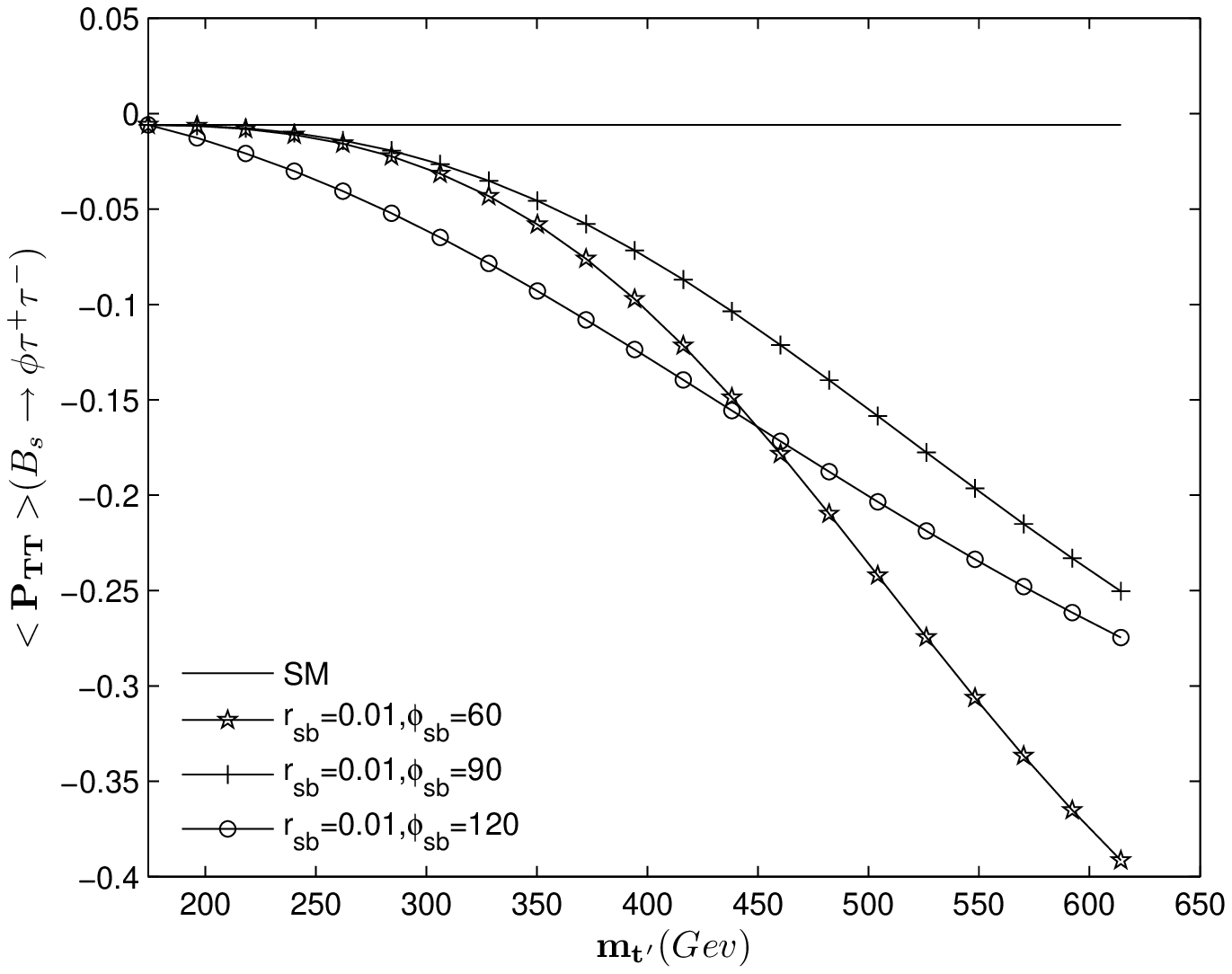}
        \includegraphics[height=2.1in]{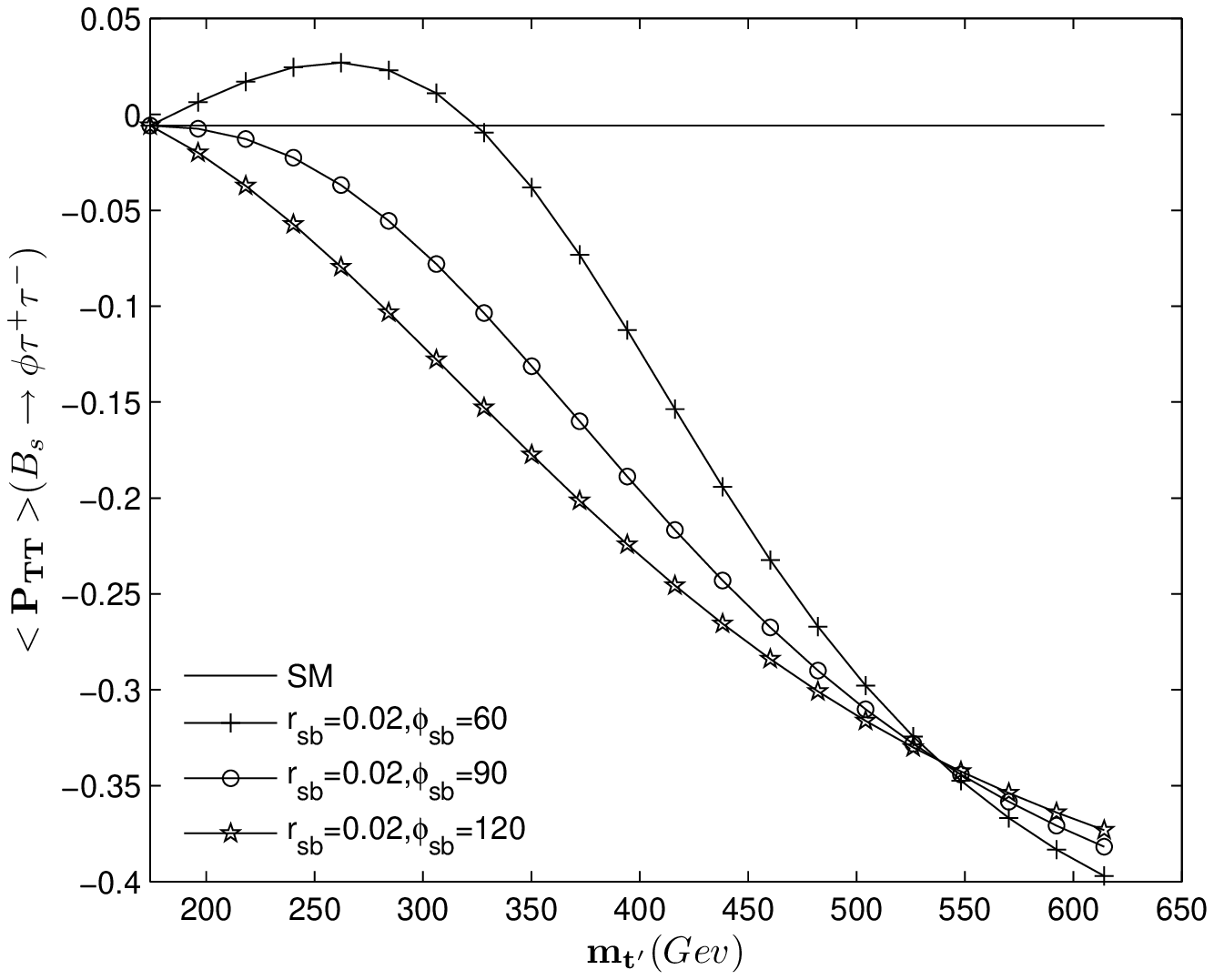}
        \includegraphics[height=2.1in]{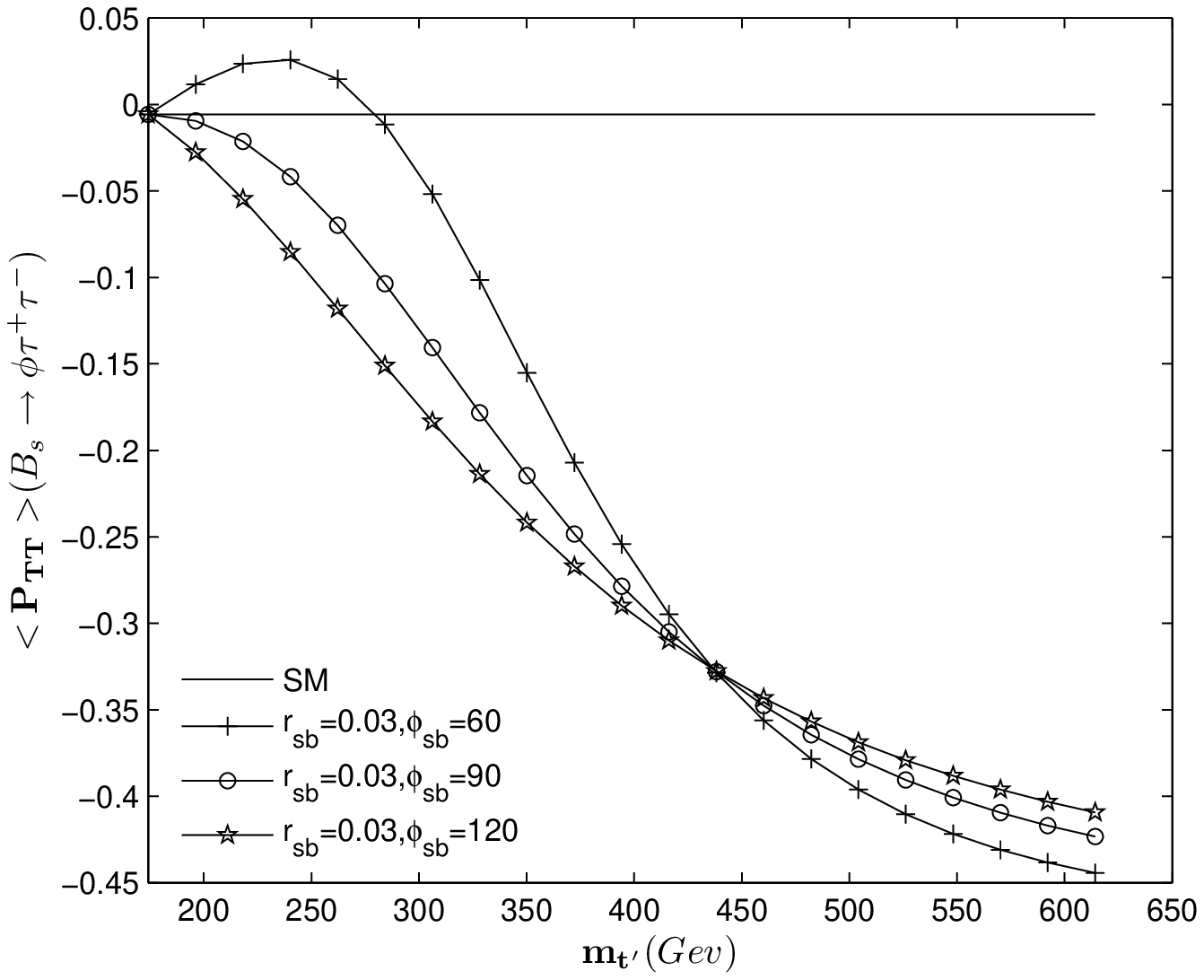}
                \caption{The dependence of the $\lla P_{TT}\rra$  on the fourth-generation quark mass
$m_{t'}$ for three different values of
 $\phi_{sb}=\{60^\circ,~ 90^\circ, ~120^\circ\}$ and $r_{sb}=\{0.01,~0.02,~0.03\}$ for the $\mu$ and $\tau$ channels.}
\label{PTT}
                  \end{minipage} }

  \end{figure}

\end{document}